\documentclass[twocolumn]{aastex62}

\newcommand{\nnb}[1]{\textcolor{black}{{#1}}}
\newcommand{\mike}[1]{\textcolor{black}{{#1}}}
\newcommand{\nna}[1]{\textcolor{black}{{#1}}}   
\newcommand{\nnd}[1]{\textcolor{black}{{#1}}}
\newcommand{\nnz}[1]{\textcolor{black}{{#1}}}
\newcommand{\nnc}[1]{\textcolor{black}{{#1}}}

\received{\today}

\shorttitle{Flare reconnection-driven magnetic field and Lorentz force variations}
\shortauthors{Barczynski et al.}
\watermark{DRAFT}

\usepackage{bm}
\usepackage{graphicx}
\graphicspath{{./}{figs-ps/}{figs/}}
\usepackage{natbib}
\usepackage{amsmath}
\usepackage{esint}
\usepackage{hyperref}
\hypersetup{
    colorlinks=true,
    linkcolor=blue,
    filecolor=magenta,      
    urlcolor=cyan,
}
\usepackage{amsmath}

\begin{document}

\title{FLARE RECONNECTION-DRIVEN MAGNETIC FIELD AND LORENTZ FORCE VARIATIONS AT THE SUN'S SURFACE.}

\correspondingauthor{Krzysztof Barczynski}
\email{krzysztof.barczynski@obspm.fr}

\author[0000-0000-0000-0000]{Krzysztof Barczynski}
\affil{LESIA, Observatoire de Paris, Universit\'e PSL , CNRS, Sorbonne Universit\'e, Universit\'e Paris-Diderot, 5 place Jules Janssen, 92190 Meudon, France}

\author[0000-0000-0000-0000]{Guillaume Aulanier}
\affil{LESIA, Observatoire de Paris, Universit\'e PSL , CNRS, Sorbonne Universit\'e, Universit\'e Paris-Diderot, 5 place Jules Janssen, 92190 Meudon, France}

\author[0000-0000-0000-0000]{Sophie Masson}
\affil{LESIA, Observatoire de Paris, Universit\'e PSL , CNRS, Sorbonne Universit\'e, Universit\'e Paris-Diderot, 5 place Jules Janssen, 92190 Meudon, France}

\author[0000-0000-0000-0000]{Michael S. Wheatland}
\affil{Sydney Institute for Astronomy, School of Physics, University of Sydney, NSW 2006, Australia}



\begin{abstract}

\nna{During eruptive flares, vector magnetograms show increasing horizontal magnetic field and downward Lorentz force in the Sun's photosphere around the polarity-inversion line. Such behavior} has often been associated with the implosion conjecture and interpreted as the result of either momentum conservation while the eruption moves upward, or of the contraction of flare loops. \nna{We characterize the physical origin of these observed behaviors by analyzing a generic 3D MHD simulation of an eruptive flare. Even though the simulation was undesigned to recover the magnetic field and Lorentz force properties,} it is fully consistent with them, and it provides key additional informations to understand them. The area where the magnetic field increases gradually develops between current ribbons, which spread away from each other and are connected to the coronal region. This area is merely the footprint of the coronal post-flare loops, whose contraction increases their shear field component and the magnetic energy density in line with the ideal induction equation. \nna{For simulated data, we computed the Lorentz force density map by applying the method used in observations. We obtained increase of the downward component of the Lorentz force density around the PIL --consistent with observations. However, this significantly differs from the Lorentz force density maps obtained directly from the 3D magnetic field and current. These results altogether question previous interpretations based on the implosion conjecture and momentum conservation with the CME, and rather imply that the observed increases in photospheric horizontal magnetic fields result from the reconnection-driven contraction of sheared flare-loops.}

\end{abstract}


\section{Introduction} \label{sec:intro}
Solar flares are the most dynamic and energetic phenomena in the solar atmosphere, often related to coronal mass ejections (CMEs, e.g., \citet{2011ApJ...738..167S, 2006SSRv..123..251F}).
Therefore, they have a dominant influence on space weather.
Hence, the investigation of flare physics plays a critical role in understanding space weather and preparing its forecast.

Solar flares were first observed by \citet{1859MNRAS..20...13C}.
Ground-based H$\alpha$ and magnetic field observations, together with sounding rocket and space-based mission data allowed the construction of models for the solar flares.
The 2D standard model named CSHKP  (after \citet{1964NASSP..50..451C, 1966Natur.211..695S, 1974SoPh...34..323H, 1976SoPh...50...85K}) explained the solar flare as caused by coronal reconnection between two opposite magnetic fields, resulting in the release of a large amount of energy and the creation of post-flare loops.

\mike{Photospheric} vector magnetic field observations are used to study the spatiotemporal evolution of the current, magnetic field and the Lorentz force.
The previous analysis shows characteristic trends in their evolution.
Here, we briefly discuss the most common trends.

The map of the photospheric electric current shows two J-shaped current ribbons located on both sides of the polarity-inversion-line (PIL).
These current ribbons characterize an opposite sign of the vertical current density ($j_{z}$) \citep{2014ApJ...788...60J}.
During the eruptive phase, the current ribbons spread outwards from each other and at the same time the current density increases in the current ribbons almost twice as compared to the pre-flare phase \citep{2014ApJ...788...60J, 2016A&A...591A.141J}.
The analysis of $j_{z}$ dynamics around the PIL shows that $j_{z}$ increases steadily until the flare occurs, then steadily decreases for the next several hours \citep{2013SoPh..287..415P}.
Moreover, $j_{z}$ decreases in the whole active region after the flare eruption \citep{2017ApJ...839..128W}.
However, \citet{2012ApJ...759...50P} notice a lack of general trends in the current density evolution around the PIL.
%

The horizontal magnetic field ($B_{\mathrm{h}}$) and the shear tend to increase around the PIL.
This is related to a decrease of the magnetic field line inclination with respect to the solar surface \citep{2010ApJ...716L.195W, 2012ApJ...745L...4L, 2012ApJ...745L..17W,  2012ApJ...759...50P, 2012ApJ...749...85G,  2014ApJ...795..128L, 2015ApJ...806..173B}.
The component of the horizontal magnetic field parallel to the PIL also increases during the flare, while the perpendicular component of $B_{\mathrm{h}}$ undergoes only minor changes \citep{2012ApJ...748...77S, 2013SoPh..287..415P}.
%
%
On the contrary, \citet{2013SoPh..287..415P} and \citet{2017ApJ...839...67S} suggest that the vertical magnetic field around the PIL is almost constant with time.
Moreover, the chromospheric line-of-sight magnetic field ($B_{\mathrm{los}}$) obtained for the first time by \citet{2017ApJ...834...26K} shows that the chromospheric magnetic field decreases nearby the PIL.

\nna{The Lorentz force density ($\bm{j \times B}$) cannot be obtained directly from the observations.
The measurement of the photospheric vector magnetic field allows us to estimate the total Lorentz force (hereafter called the alternative Lorentz force, $\bm{F}$) on the coronal volume as the surface integral coming from the volume integral of the Maxwell stress tensor \citep{2012ApJ...759...50P, 2012SoPh..277...59F}. \nnz{The maps of the changes of the photospheric magnetic field can be used to calculate the total Lorentz force change in the corona and on the whole photospheric plane \citep{2017ApJ...839...67S}.} However, to create maps of Lorentz force, it is common to use the sole integrand of $\bm{F}$ (thereafter the alternative Lorentz force density, $\bm{s}^\mathrm{ad}$) \citep{2012ApJ...759...50P}.}
\nna{The increase of the negative vertical $\bm{s}^\mathrm{ad}$ in the solar corona is co-spatial and co-temporal with the magnetic field increase \citep{2010ApJ...716L.195W, 2011ApJ...727L..19L, 2012ApJ...759...50P, Sarkar2018}.}
Moreover, the horizontal Lorentz force starts to be more parallel to the PIL and acts in opposite directions on each side of the PIL \citep{2012ApJ...759...50P}.
\citet{2011ApJ...727L..19L} and \citet{2012ApJ...759...50P} suggest that the trends described above are consistent with the magnetic field implosion \nna{scenario} proposed by \citet{2008ASPC..383..221H}. 
This naturally raises a question: is the magnetic field implosion responsible for the increase of the horizontal magnetic field around the PIL?

The joint observed increases in horizontal fields and in downward \nna{alternative} Lorentz force density on the photosphere have been qualitatively interpreted as a back-reaction on the photosphere of the eruption. Two different specific mechanisms have been proposed.
The first idea is the downward acceleration of the photospheric plasma itself, as a response to a disruption of the force balance in the corona \citep{2008ASPC..383..221H}, in particular associated with momentum conservation as the eruption is accelerated upwards \citep{2011SSRv..158....5H, 2012SoPh..277...59F}.
The second idea is the collapse of sheared loops from the corona above the photosphere \citep{2012ApJ...759...50P, 2012ApJ...748...77S}, in particular the post-reconnection loops above the PIL \citep{2011ApJ...727L..19L} that form according to the CSHKP model.
These two ideas were often related to the implosion conjecture, that associates local magnetic energy release during flares and/or CMEs with local volume decrease \citep{2000ApJ...531L..75H, 2015A&A...581A...8R}.
However, \citet{2017ApJ...839...67S} questioned its application to the photosphere.

\nnb{Based on the 3D MHD model of an eruptive flare, we try to understand why the horizontal magnetic field increases around the PIL.}
\nnb{To this aim,}  we consider the following questions: what is a reason for the current density and the magnetic field evolution during the flare?
Can the \nna{alternative Lorentz force density}
coming from the volume integral of the Maxwell stress tensor (\nna{$\bm{s}^\mathrm{ad}$}) be used as a proxy of the Lorentz force density ($\bm{j} \times \bm{B}$)?
What determines the temporal variability of the horizontal magnetic field?

This paper is divided into five sections. 
Section~\ref{sec:mhd} presents a short description of the \nna{3D MHD simulation used to  model an eruptive flare.}
Section~\ref{sec:mf_cur_var} describes the \nna{evolution} of the current and magnetic field at the photosphere during the eruption phase.
In Section~\ref{sec:lorentz_force_comp} we report the difference between the Lorentz force density \nnb{($\bm{j}\bm{\times} \bm{B}$)} and \nna{the alternative Lorentz force density ($\bm{s}^\mathrm{ad}$)}.
Section~\ref{sec:induction_equation} reports the \nna{evolution} of the horizontal magnetic field as implied by the induction equation. 
We conclude our analysis in Section~\ref{sec:sumcon}.


\section{MHD model of eruptive events}\label{sec:mhd} 

We investigate the results of an eruptive flare simulation provided by \citet{2015ApJ...814..126Z}.
This simulation presents the solution of the zero-$\beta$ (pressureless), time-dependent MHD equations with the OHM-MPI code \citep{2005A&A...430.1067A}.
The dense photosphere is approximated as a line-tied boundary at $z=0$.
This common approximation is roughly valid during the time of eruption albeit for some small leaks (see, \citet{2008A&A...490..353G} and \citet{2018ApJ...864..159W}).
We analyze two large-scale patches of opposite magnetic field polarities based on the simulation labelled as ``Run D2''.

This paper focuses on the eruptive phase of the flare, showing the flare evolution between $t_{\mathrm{0}}=164t_{\mathrm{A}}$ at which the torus instability occurs and $t_{\mathrm{end}}= 244t_{\mathrm{A}}$ at which the simulation breaks due to unresolved gradients in the coronal flare current sheet.
The simulation uses a non-uniform mesh ($n_{x} \bm{\times} n_{y} \bm{\times} n_{z}=251 \bm{\times} 251 \bm{\times} 231$) with a domain of $x,y\in [-10; 10]$ and $z\in[0; 30]$.
 However, our analysis is limited to the flare region of $x\in[-3;  2]$,  $y\in[-4.5; 3]$, and $z\in[0; 5]$.
 %
 
 The model provides the spatiotemporal evolution of the component of the vector magnetic field ($\bm{B}$), electric current-density ($\bm{j}$), and plasma velocity ($\bm{u}$).
 More details of the model and ``Run D2'' simulation are discussed in \citet{2015ApJ...814..126Z}.

\begin{figure*}[ht!]
\epsscale{1.2}
\plotone{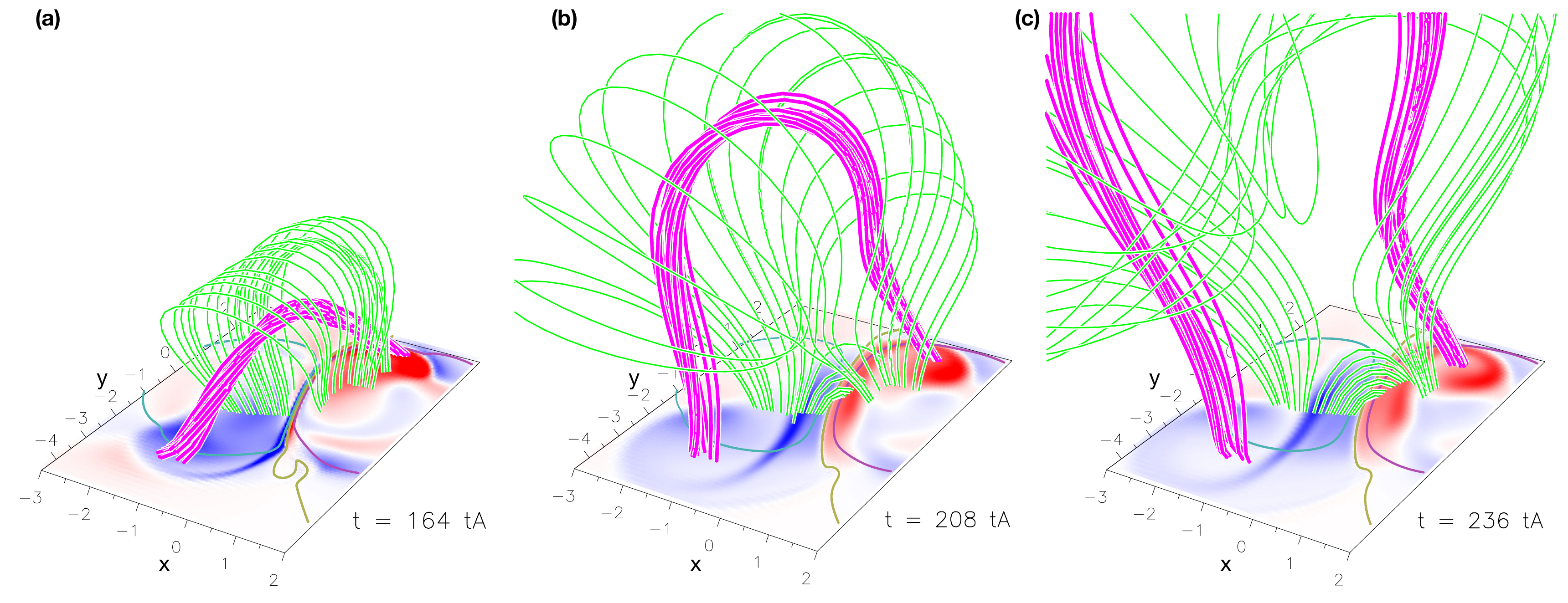}
\caption{Three-dimensional plots of the coronal magnetic field line evolution showing the underlying photospheric currents. The snapshots present projected views of the flare simulation before reconnection ($t=164t_{\mathrm{A}}$), during the eruption ($t=208t_{\mathrm{A}}$), and at a later time ($t=236t_{\mathrm{A}}$). The blue/red color scale images indicate the negative/positive component of the vertical current density $j_{z}$. Colors are consistent with the currents presented in the next figures. Light blue and purple contours define the two opposite magnetic field patches at the photosphere. The polarity inversion line is highlighted in olive color. The magenta line shows the magnetic flux rope rooted within the current ribbons' hook. The bright-green lines present the overlying arcade. \label{fig:f1}}
\end{figure*}

\nnz{The output of the simulation is presented in a dimensionless form, where the magnetic permeability is set to $\mu=1$  \citep{2010ApJ...708..314A, 2015ApJ...814..126Z}. 
The time unit $t_A = 1$ is the transit-time of an Alfven wave from the PIL to the center of one polarity at $z=t=0$. The spatial unit $L=1$ is the distance between the PIL and the center of one polarity (see e.g. Figures~\ref{fig:f1} and~\ref{fig:f2}). The magnetic field unit is arbitrary, and  $B_z(z=0.1)$ peaks at 3.26. The simulated dimensionless value can be dimensionalizing to physical units: for a young emerging active region, $B_z$ peaks at 2000 G and the distance between two polarities is about $2*L = 20 Mm$. This leads to a physical normalization of $B_{z \circ} = 2000/3.26 =  613 G$ and $L_{\circ} = 20/2 = 10 Mm$. The example of dimensionalizing is presented in Section~\ref{subsec:b_para}.}


Figure~\ref{fig:f1} presents the evolution of magnetic field lines in the corona and electric currents right above the line-tied photospheric boundary.
This figure shows the area of two large-scale opposite magnetic field polarities and their surroundings, \nnz{centered} at $[x;y]=[-1; -1]$ and $[x;y]=[1; 1]$ for the negative (light blue contour) and the positive (pink contour) polarity respectively.
Using the TOPOlogy and field line TRacing code \citep[TOPOTR;][]{1996JGR...101.7631D} we plotted magnetic field lines to show and interpret the dynamics of the eruption.

Before the eruption (Figure~\ref{fig:f1}a), the electric currents are concentrated in two J-shaped ribbons (red/blue color concentration) located close to the polarity-inversion-line (PIL, olive color).
More detailed analysis of the currents is presented in Section~\ref{subsec:tem_evol_current}.
The magnetic field lines (green), rooted along the straight segment of the PIL, create the $\Omega$-shaped arcade connecting the two opposite polarities.
The magnetic field of these field lines is almost vertical near the surface.
 The magnetic flux rope (purple) is rooted inside the current ribbon hooks ($[x;y]\approx[-1; 2.5]$ and $[x; y]\approx[0.5; 1.5]$) and is overlaid by the $\Omega$-shaped arcade.
Initially, the flux rope also has an arched shape and a low-inclination with respect to the local solar surface.
At around $t_{\mathrm{crit}}=165 t_{\mathrm{A}}$, the flux rope becomes torus unstable and erupts \citep{2015ApJ...814..126Z, 2017ApJ...837..115Z}. 
%

The layers of strong gradients of connectivity of the magnetic field are defined as the quasi-separatrix layers (QSLs; \citet{1996A&A...308..643D, 1997A&A...325..305D}).
In this domain, high-current density layers are created, whose photospheric footprints are the current ribbons \citep{2005A&A...444..961A, 2013A&A...555A..77J}.
The core of the QSL is the hyperbolic flux tube (HFT, \citet{2002JGRA..107.1164T}).
During the flare evolution, the current layer becomes thinner, especially around the HFT.
When the current layer is thin enough, the reconnection begins there \citep{2006SoPh..238..347A}.
As the result of the reconnection (Figure~\ref{fig:f1}b), the new post-flare magnetic field lines (green) close down above the PIL.
These lines have a small angle between the field lines and the solar surface.
Moreover, they are significantly shorter than the pre-flare magnetic field lines.
At the same time, the flux rope (purple lines) rises strongly and its magnetic field at the solar surface becomes more vertical.

Subsequently, the simulation shows a fast outward expansion of the overlying magnetic arcade (green lines) which reconnects at the HFT, building up further the flux rope.
The upper part of the flux rope expands upward, while the middle part moves towards the current sheet.
Therefore, the shape of the flux rope becomes more complex (see Figure 1 in \citet{2017ApJ...844...54D}).
Moreover, the comparison of the magnetic field lines (green) of the $\Omega$-shaped arcade in Figure~\ref{fig:f1}a and Figure~\ref{fig:f1}b clearly shows that the magnetic field at the negative part of the $x$-axis is more vertical than at $x>0$, where the magnetic field is significantly more horizontal.
The flux tube asymmetry results from the asymmetric magnetic field configuration in the simulation initialization and is responsible for the CME deflection towards the negative $x$-axis.


\section{\nna{The electric current density and the magnetic field}} \label{sec:mf_cur_var} 

\begin{figure*}[ht!]
\epsscale{1.2}
\plotone{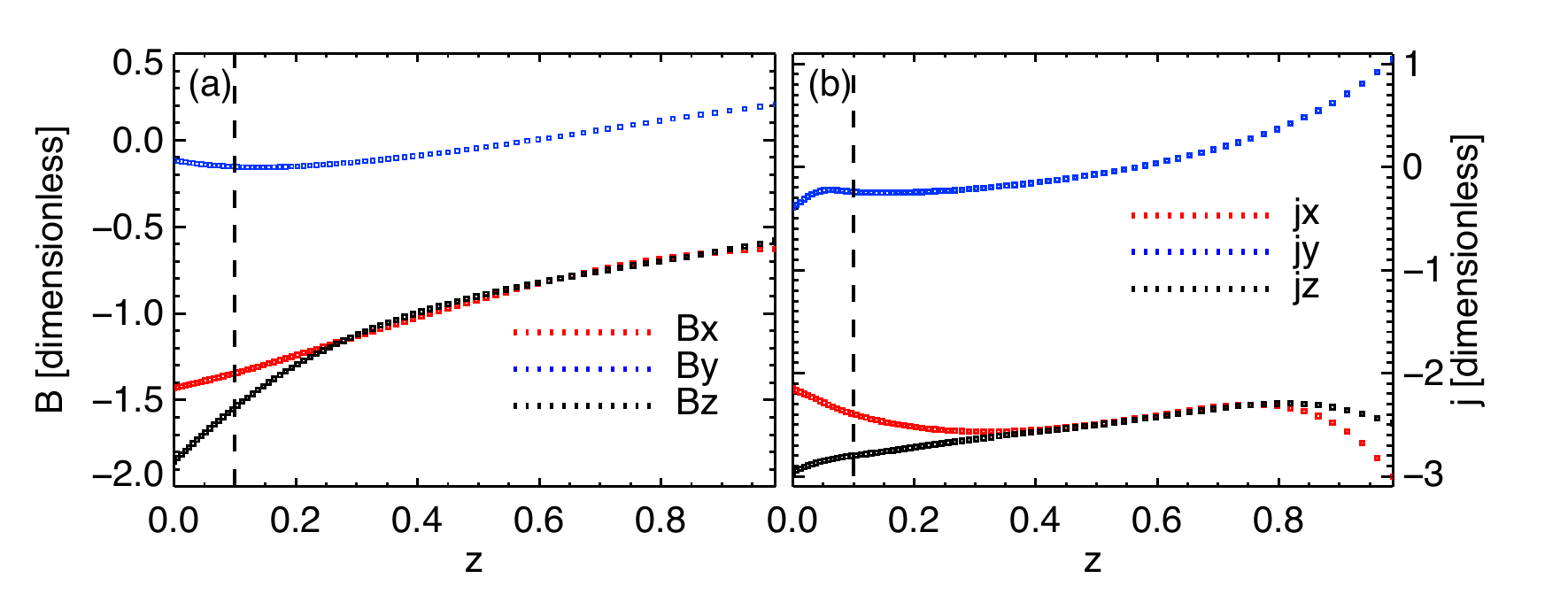}
\caption{\nnz{The evolution of magnetic field (a) and electric current density (b) with height on the solar atmosphere from $z=0.0$ to $z=1.0$. The plots present the vector components (color-coded squares) during the eruption ($t=208t_{\mathrm{A}}$) at $[x; y]=[-0.6; 0.2]$. The dash line marked $z=0.1$.}\label{fig:f02}}
\end{figure*}

Observations of the solar photosphere allow us to study the spatiotemporal variability of the 2D-distributions of the magnetic field and electric current density at the solar surface (e.g., \citet{2012ApJ...759...50P}).
We analyze our 3D MHD simulation \mike{in the same way, i.e., using the photospheric magnetic field to derive the current density.}
The analysis is carried out on the ($x$, $y$) plane, parallel to the photosphere at $z=0.1$. 
\nnz{The influence of the boundary layers is not obvious for $B(z)$ (Figure~\ref{fig:f02}a). However a small boundary layer is visible for $j(z)$, especially for $j_y$, for $z<0.05$ (Figure~\ref{fig:f02}b).
Therefore, we chose the layer at $z=0.1$ to avoid the influence of boundary effects seen at lower $z$.}
First, a very low lying bald-patch separatrix stands right on the PIL, it is the site of the currents along the PIL not relevant for this study \citep{2019A&A...621A..72A}. We focus on an altitude above the bald-patch separatrix.
Second, a boundary layer exists at lower $z$, due to an artificially prescribed diminishing of the Lorentz forces right above the boundary during the eruption, which is applied to prevent numerical instabilities \citep{2015ApJ...814..126Z}.
This prescription in turn leads to relatively noisy and unreliable Lorentz forces for $z<0.1$.
Results are shown at $z=0.1$ in the different figures in this paper, but the patterns are not specific to this altitude and can seen down to $z=0$ and for $z\gtrsim0$.
%

\subsection{The electric current density} \label{subsec:tem_evol_current}

Using the modeled vector magnetic field ($\bm{B}$) from the simulation, we calculate the current density vector ($\bm{j}$).
To this purpose, we use Amp\`ere's circuital law (Equations~\ref{eq:ampere_law}): 
\begin{equation} \label{eq:ampere_law}
\bm{\nabla} \bm{\times} \bm{B} =\mu_{0}\bm{j},
\end{equation} 
and use the centered difference method for each mesh point.

\begin{figure*}[ht!]
\epsscale{1.2}
\plotone{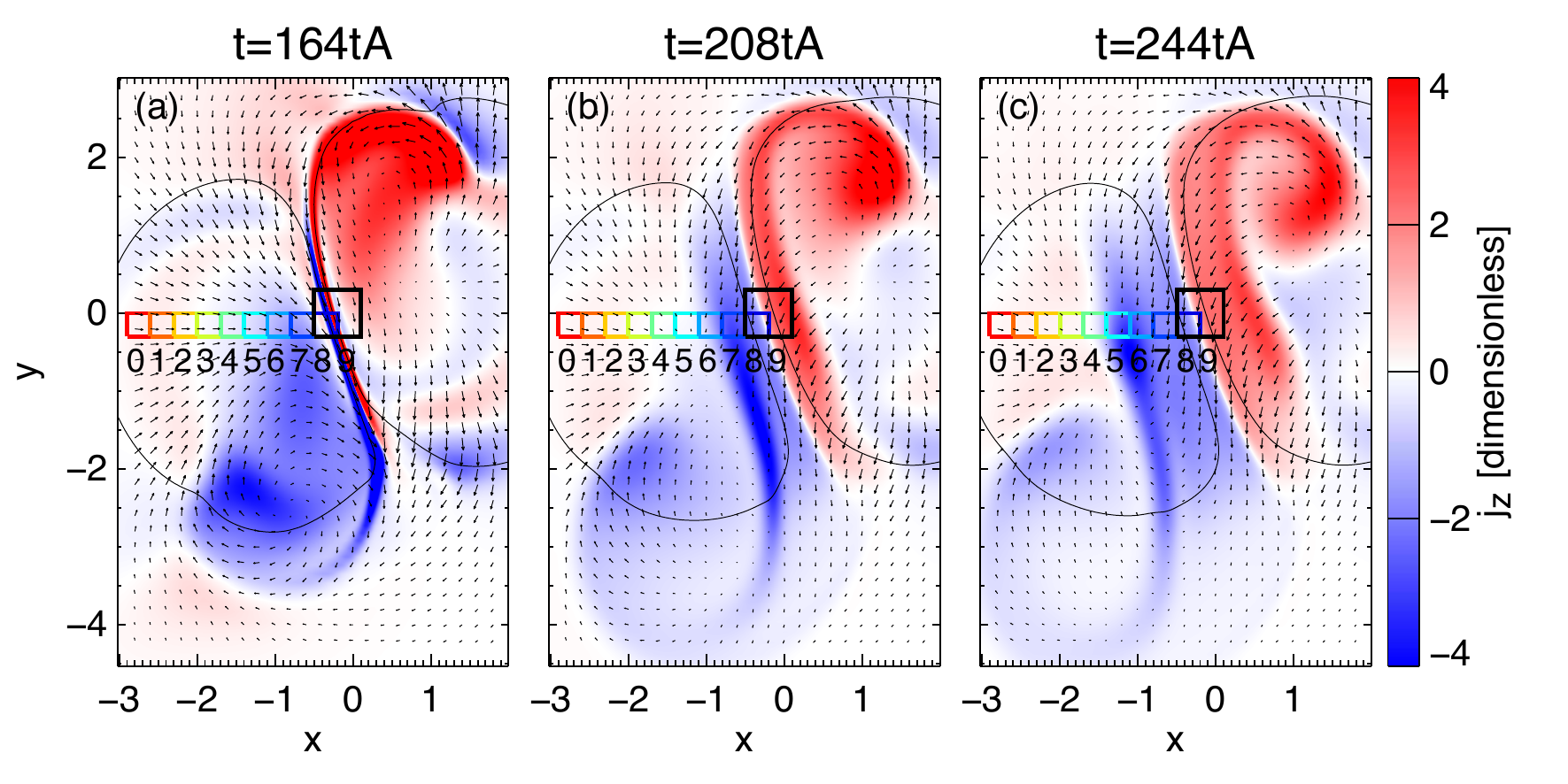}
\caption{The vertical component of the current density $j_z(z=0.1)$ at three different times during the eruption. The negative/positive currents are defined by the blue/red color scale. Arrows represent magnetic field components $B_{x}(z=0.1)$ and $B_{y}(z=0.1)$. The black contours indicate the vertical component of magnetic field \nnz{$B_{z}(z=0.1)=-1$ and $B_{z}(z=0.1)=1$}. The color coded-boxes (0-9) mark regions that are used in further analysis. 
\\
(An \href{run:./ma.mp4}{animation}  of this figure is available in the online journal.) \label{fig:f2}}
\end{figure*}

First, we consider the spatiotemporal evolution of the vertical component of the electric current, $j_{z}$, which is presented in Figure~\ref{fig:f2}.
The spatial distribution of $j_{z}$ clearly shows positive and negative (red and blue) J-shaped electric current ribbons (in almost the entire field-of-view (FOV)) with a clear hook at the end of each ribbon ($[x;y]\approx[-1; 2.5]$ and $[x; y]\approx[0.5; 1.5]$).
The straight part of the current ribbons ($x\in[-1;1]$ and $y\in[-2.5; 2]$) is almost symmetric with respect to the PIL. 
%
%
Moreover, large-scale $j_{z}$ concentrations also exist far away from the PIL (e.g., $[x;y]\approx[-1.5; 2]$).
These structures form complex and irregular patches of significantly lower $j_{z}$ than in the current ribbons.

To analyze the temporal evolution of the $j_{z}$ distribution, we compare maps at the start (Figure~\ref{fig:f2}a, $t=164t_{\mathrm{A}}$), middle (Figure~\ref{fig:f2}b, $t=208t_{\mathrm{A}}$), and the end (Figure~\ref{fig:f2}c, $t=244t_{\mathrm{A}}$) of the simulation.
Two opposite-sign electric current ribbons are visible during the whole eruption phase.
Initially (Figure~\ref{fig:f2}a), the current ribbons are almost symmetric with respect to the PIL.
Nevertheless, this symmetry is broken during the flare eruption (Figure~\ref{fig:f2}b) and is caused by the system deflection (Section~\ref{sec:mhd}).
The comparison of $j_{z}$ maps at $t=164t_{\mathrm{A}}$ (Figure~\ref{fig:f2}a) and $t=208t_{\mathrm{A}}$ (Figure~\ref{fig:f2}b) indicates that the current ribbons are moving away from each other, thus also from the PIL, along the $x$-axis towards the centers of the magnetic field concentrations.
This is accompanied by the strong decrease of $j_{z}$ at the current ribbons and their broadening (Figure~\ref{fig:f2}c).
The current ribbon hooks also evolve with time.
Initially (Figure~\ref{fig:f2}a), their shape is similar to the end of the letter ``J'', but in the final phase of the simulation, those hooks become rounder and almost closed (Figure~\ref{fig:f2}c), more like a ``$\sigma$''.

\subsection{The horizontal magnetic field} \label{subsec:ch_bh}

We investigate the photospheric horizontal magnetic field changes,
\begin{equation} \label{eq:bxych} 
\delta B_{\mathrm{h}}=B_{\mathrm{h}}(\delta t)-B_{\mathrm{h}}(t_{0})~\mathrm{with}~B_{\mathrm{h}}=\sqrt{B^{2}_{x}+B^{2}_{y}}~,
\end{equation} 
where $\delta t$ is the simulation time of the analyzed frame, and  $t_{\mathrm{0}}=164t_{\mathrm{A}}$ is the reference frame's time.

\begin{figure*}[ht!]
\epsscale{1.2}
\plotone{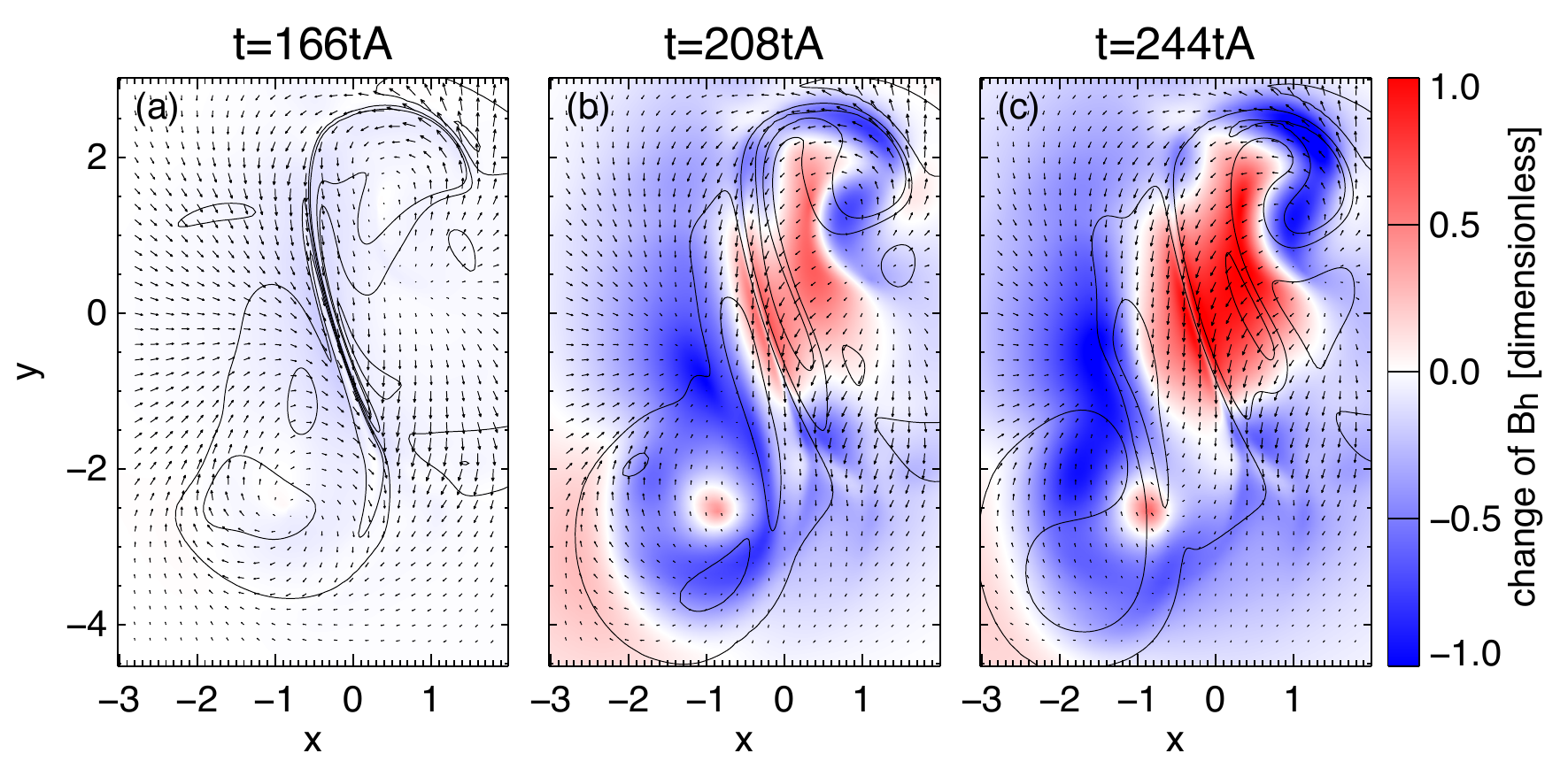}
\caption{\nnc{The magnitude of the horizontal magnetic field changes at three different times during the eruption. The change of the absolute value of the horizontal magnetic field $\delta B_h(z=0.1)$ is defined by the color-code. The arrows are the same as in Figure~\ref{fig:f2}. The solid contour lines mark electric current density at levels $j_{z}(z=0.1)$=-2.2; -0.4; 1.6 and 2.7 (see $j_z$ maps in Figure~\ref{fig:f2})}. 
\\
(An \href{run:./mb.mp4}{animation}  of this figure is available in the online journal.)
\label{fig:f3}}
\end{figure*}

The spatiotemporal evolution of $\delta B_{\mathrm{h}}$ is presented in Figure~\ref{fig:f3}. 
These maps highlight the significant increase of $B_{\mathrm{h}}$ around the straight part of the current ribbons ($x\in[-1;1]$ and $y\in[-2.5; 2]$).
This increase is asymmetric with respect to the PIL.
The asymmetry is stronger at the middle phase of the eruptive flare (Figure~\ref{fig:f3}b) and slightly reduced at the end of the simulation (Figure~\ref{fig:f3}c).
$B_{\mathrm{h}}$ strongly decreases at the center of the magnetic field polarities and at the end of the hooks.
In the rest of the map, $B_{\mathrm{h}}$ slightly decreases (Figure~\ref{fig:f3}b, c).
The trends described above are steady until the end of the simulation (Figure~\ref{fig:f3}c).  

The spatial relation between $\delta B_{\mathrm{h}}$ (Figure~\ref{fig:f3}) and the evolution of the vertical component of the electric current in the ribbons (dashed contours) is complex.
The region where $B_{\mathrm{h}}$ increases ($x\in[-1.5; 0.5]$ and $y\in[-2; 2]$) \mike{spreads} away from the PIL, similar to the current ribbon (Figure~\ref{fig:f2}).
On the left side of the PIL ($x<0$), the straight part of the current ribbon and the region of increasing $B_{\mathrm{h}}$ spread along the negative direction of the $x$-axis, however, the current ribbon spreads away from the PIL faster than $\delta B_{\mathrm{h}}$ as shown explicitly in Section~\ref{subsec:b_para}.
On the right side of the PIL ($x>0$), $\delta B_{\mathrm{h}}$ and $j_{z}$ spread along the positive direction of the $x$-axis and the sweep of the current ribbon precedes the high-increase of $B_{\mathrm{h}}$.
%

\begin{figure*}[ht!]
\epsscale{1.2}
\plotone{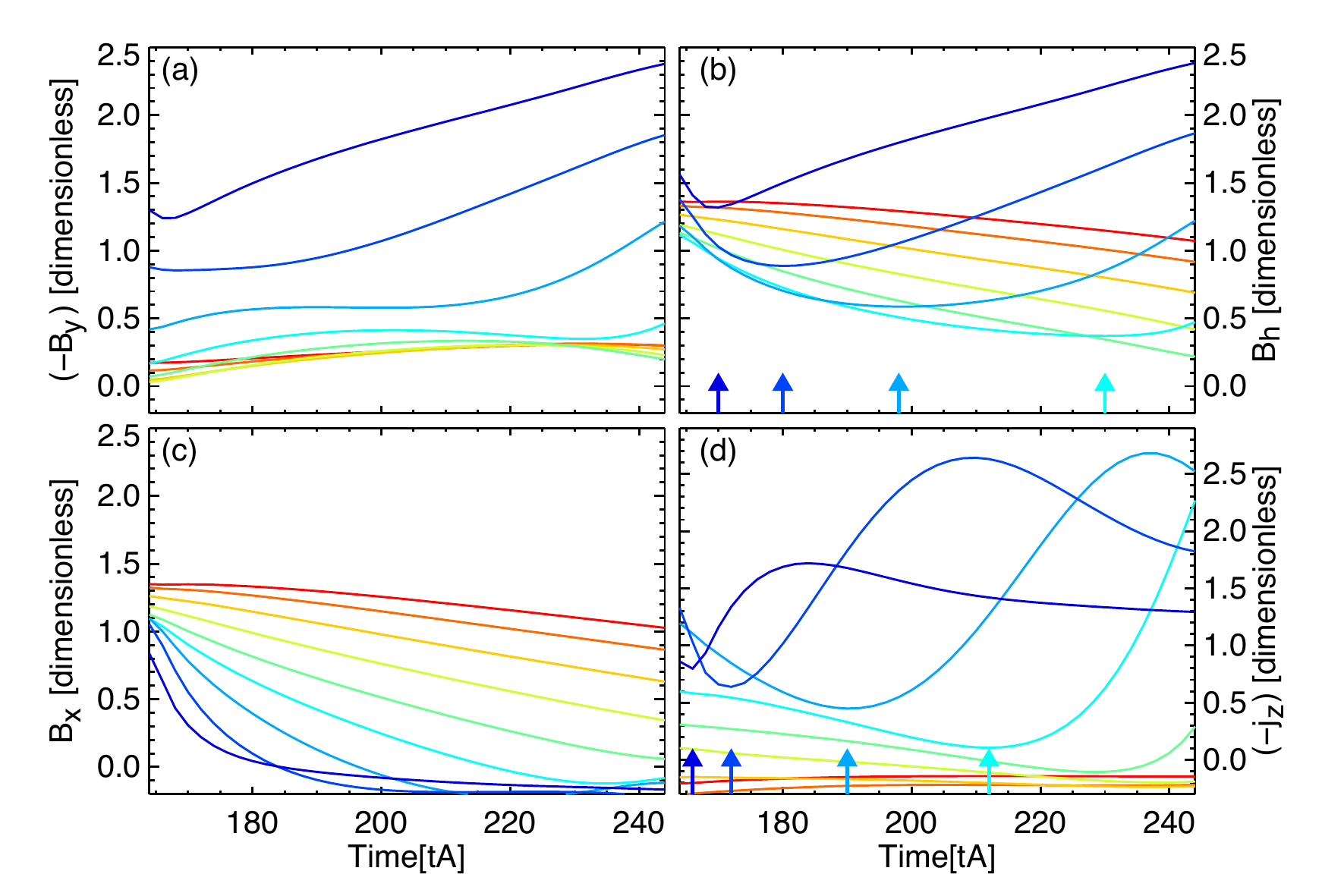}
\caption{The evolution of magnetic field and electric current density. The evolution of magnetic field components: $B_{y}$(a), $B_{\mathrm{h}}$(b), $B_{x}$(c) and current density $-j_{z}(d)$ from $t=164t_{\mathrm{A}}$ to $t=244t_{\mathrm{A}}$. The color-coded lines indicated regions-of-interest (boxes 0-9), in Figure~\ref{fig:f2}, in which these quantities are calculated. The same colors were used to identify the arrows that mark the minima of $B_{\mathrm{h}}$ and $-j_{z}$. \label{fig:f4}}
\end{figure*}

The analysis of $B_{\mathrm{h}}$ (vectors) around the PIL suggests that the $B_{\mathrm{h}}$ increase corresponds to a growth of the $B_{y}$ component.
To understand the evolution of the magnetic field nearby the PIL, more detailed analysis is required.
Thus, we choose ten regions-of-interest (ROIs) marked by color-coded boxes (Figure~\ref{fig:f2}).
We analyze separately the ROIs which are unswept by the current ribbon (ROIs 0-4), and which are swept by the current ribbon (ROIs 5-8), and ROI-9 which is located at the PIL.
%
%
To avoid the influence of the flux rope deflection \nnb{(e.g., additional $B_{x}$ increase)}, we locate ROIs on the left side of the PIL.
In each ROI, we calculate the average $B_{x}$, $B_{y}$, $B_{\mathrm{h}}$ and show their temporal evolution in Figure~\ref{fig:f4}a-c.
These average values are computed as a weighted arithmetic mean, where the non-uniform-mesh area is used as the weight.

First, we focus on ROIs which are unswept by the current ribbon (ROI 0-4, yellowish and reddish lines). 
In this case, $B_{\mathrm{h}}$ decreases (Figure~\ref{fig:f4}b) with simulation time.
This is caused by the decrease of the $B_{x}$ component (Figure~\ref{fig:f4}c).
The negative $B_{y}$ (Figure~\ref{fig:f4}a) presents only a minor increase.
%

Second, we describe the magnetic field evolution of ROIs which are swept by the current ribbon (ROI 5-8, bluish lines).
Before these ROIs are swept by the current ribbon (e.g., at $t<198t_{\mathrm{A}}$ for ROI-6), $B_{\mathrm{h}}$ slightly decreases (Figure~\ref{fig:f4}b).
It corresponds to the decrease of $B_{x}$ (Figure~\ref{fig:f4}c), while $B_{y}$ stays almost constant at that time (Figure~\ref{fig:f4}a).
Let us now look at the situation when the ROI is swept by the current ribbon and after that (e.g., at $t\geq198t_{\mathrm{A}}$ for ROI-6).
In this case, $B_{\mathrm{h}}$ significantly increases with time (Figure~\ref{fig:f4}b).
This behavior is due to the growth of the negative component of $B_{y}$ (Figure~\ref{fig:f4}a).
Meanwhile, $B_{x}$ becomes weak (around zero), then stays almost constant (Figure~\ref{fig:f4}c).
These trends remain unchanged until the end of the simulation.
%

\begin{figure*}[ht!]
\epsscale{1.2}
\plotone{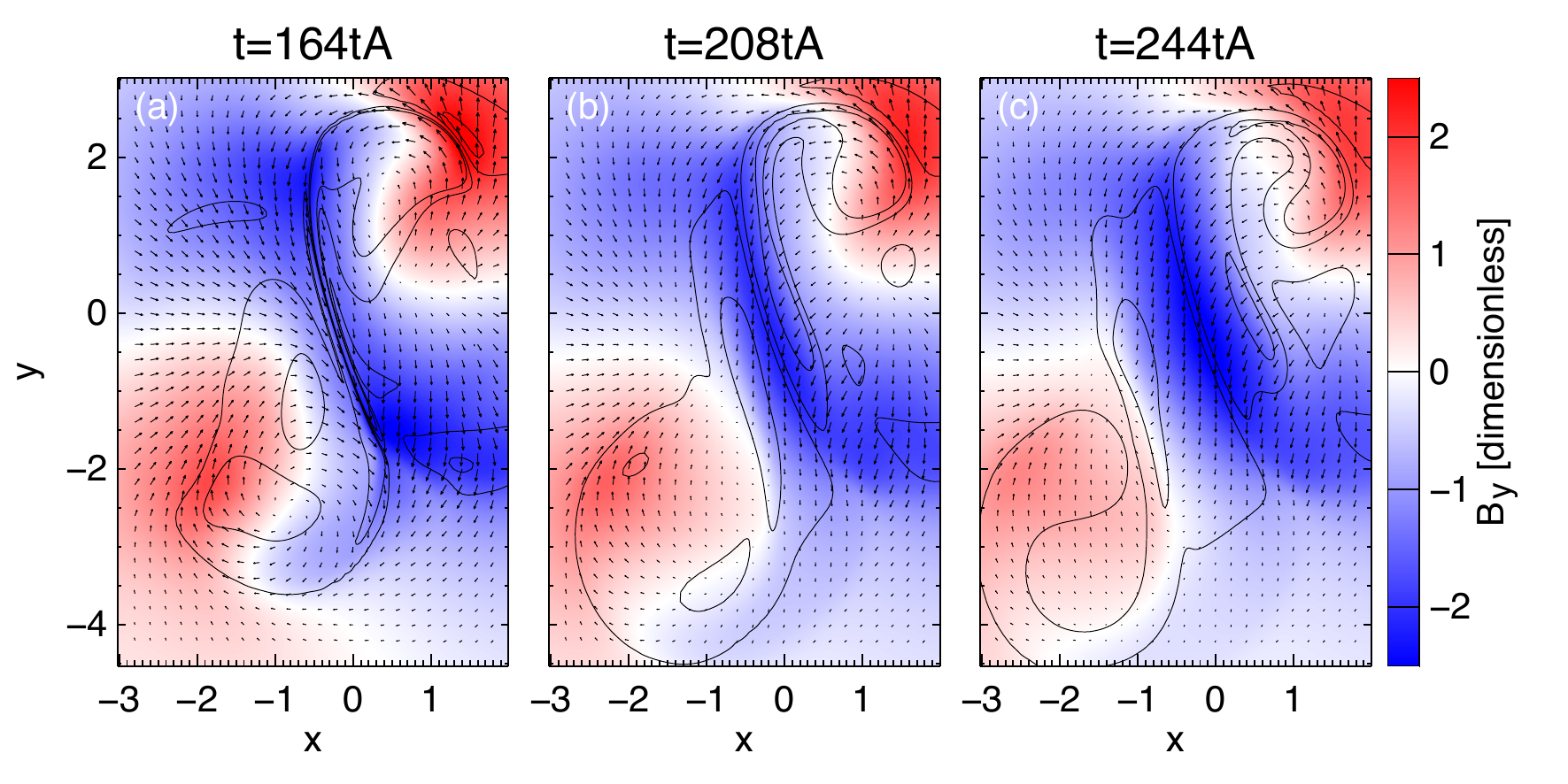}
\caption{
The horizontal magnetic field at three different times during eruption phase. Horizontal component $B_{y}$ of the magnetic field, almost parallel to the PIL, is defined by the color-code. Arrows and electric current density contours are the same as in Figure~\ref{fig:f3}. 
\\
(An \href{run:./mc.mp4}{animation}  of this figure is available in the online journal.)
\label{fig:f5}}
\end{figure*}

In ROIs which are swept by the current ribbon, the evolution of the horizontal magnetic field is determined by $B_{y}$ (Figure~\ref{fig:f4}a,b).
Here, we focus on $B_{y}$. \mike{Its} spatiotemporal evolution in the full-field-of-view is shown in Figure~\ref{fig:f5}.
$B_{y}$ is almost parallel to the PIL (see vectors).
During the eruption, the current ribbons (black contours) are moving away from each other.
Then, the (negative) $B_{y}$ component increases (in absolute value) in the region between the straight parts of the current ribbons ($x\in[1.5; 0.5]$ and $y\in[-2; 2]$).
The (negative) $B_{y}$ component decreases (in absolute value) in the rest of the map, especially nearby the hooks.
This trend lasts until the simulation ends (Figure~\ref{fig:f5}c).
 The maps of $B_{y}$ (Figure~\ref{fig:f5}) show only slight asymmetry of the magnetic field concentration around the PIL, which is the opposite to $\delta B_{\mathrm{h}}$ (Figure~\ref{fig:f3}).

\subsection{Vertical electric current density and the magnetic field dependencies} \label{subsec:b_para}

The detailed analysis of the temporal evolution of $j_{z}$ has been done for the same ROIs as the magnetic field analysis (Section~\ref{subsec:ch_bh}).
Figure~\ref{fig:f4}d shows that $-j_z$ decreases or stays almost constant in the case of the ROIs which are not swept by the current ribbon (reddish, yellowish and light-green curves, ROI 0-4).
For ROIs which are swept by the current ribbon (bluish curves, ROI 5-8), $-j_z$ initially decreases and then reaches its minimum (e.g., at $t=190t_{\mathrm{A}}$ for ROI-6). 
When the current ribbon is moving in the ROI direction, $-j_z$ significantly increases and reaches its peak at the time when the ROI is swept by the center of the current ribbon (e.g., at $t=236t_{\mathrm{A}}$ for ROI-6).
Then, $-j_{z}$ slightly decreases with the simulation time.

We analyze the temporal dependence between $B_{\mathrm{h}}$ (Figure~\ref{fig:f4}b) and negative $j_{z}$ (Figure~\ref{fig:f4}d) in ROIs where the current ribbon sweeps the ROIs.
For this purpose, we specified when the $B_{\mathrm{h}}$ and $-j_z$ curves reach a minimum, in other words, we define when $B_{\mathrm{h}}$ and $-j_z$ begin to grow.
Their minima are highlighted by arrows in Figure~\ref{fig:f4}b and d.
The comparison of $B_{\mathrm{h}}$ (Figure~\ref{fig:f4}b) and $-j_z$ (Figure~\ref{fig:f4}d) minima clearly shows that the negative $j_{z}$  always starts to grow before $B_{\mathrm{h}}$ begins to rise.
Moreover, the time differences between minima of $-j_z$ and $B_{\mathrm{h}}$ increases with the simulation time (see also Section~\ref{subsec:ch_bh}), from $4t_{\mathrm{A}}$ for the ROI-8 (dark-blue curve) to even $18t_{\mathrm{A}}$ for the ROI-5 (cyan curve).

In Section~\ref{subsec:tem_evol_current}, we present the \nna{evolution} of the electric current density.
\citet{2015ApJ...814..126Z} presented that the magnetic flux rope is formed as the result of the flux cancellation at the PIL of the previously sheared magnetic arcade.
In our simulation, the characteristic sign of a coming eruption is the flux-rope expansion, which is caused by the torus instability \citep{2010ApJ...708..314A}.
The flux rope expansion influences the $\Omega$-shaped arcade.
It causes the magnetic fields lines below the flux-rope, but still in the corona, to come closer to each other \nna{and to reconnect at the flare current sheet behind the torus unstable flux rope. This leads to the formation of flare-reconnection driven contracting loops} \citep{1997SoPh..176..153M, 2012ApJ...758...60F}.

%
In Section~\ref{subsec:tem_evol_current}, we show that, during the flare, the current ribbons spread away from each other. 
The thinning of the flare current sheet in the corona produces currents all along the quasi-separatrix layers down to their footpoints, and hence the $j_{z}$ increase at the boundary and right above it.
The spreading of the ribbons is due to the reconnections \nna{of overlying magnetic loops} one after the other that change the double-J shaped QSL positions, as explained in \citet{2013A&A...555A..77J}.

For the ROIs 5-8, the electric current increases first, reaching its peak when the ROI is swept by the current ribbon. 
This is natural because the current ribbon is the \nna{photospheric trace of the volume} electric current concentrations \nna{in the reconnecting loops}. 
Then, the current density slightly decreases in regions which were swept by the current ribbon.
\citet{2014ApJ...788...60J} presents two theoretical arguments that confirm our finding.
First, due to the eruption, the length of the pre-flare magnetic field lines increases during the flare (see the green lines in Figure~\ref{fig:f1}), but the twist end-to-end is conserved.
This implies the decrease of the current density across the flux rope. 
Second, the magnetic field is more potential after the reconnection.

Due to the eruption, the HFT is moving up, the same as the X-point moving up in the 2D model.
First, the reconnection involves the loops rooted near the PIL, and then the loops rooted further and further from the PIL.
Therefore, the distance between the foot-points of the newly formed post-flare loops increases during the simulation.
Hence, the post-flare loops created at the start of the eruption are shorter than the post-flare loops formed at the end of the simulation.
The newly formed post-flare loops are rooted at the outer edge of the current ribbon, but previously formed post-flare loops still exist where the new ones are created.
Hence, the current ribbon is wider with time (Section~\ref{subsec:tem_evol_current}). 
%

Let us now focus on the magnetic field evolution around the straight part of the current ribbons (Section~\ref{subsec:tem_evol_current}) during the flare.
The horizontal magnetic field in ROIs swept by the current ribbon first quickly decreases (e.g., for ROI-8 from $t=164t_{\mathrm{A}}$ to $t=170t_{\mathrm{A}}$), which is related to the $B_{x}$ decrease because $B_{y}$ is almost constant at this same time and place.
\nna{Before the eruption, the geometry of the loops overlying the flux rope gives a positive $B_{x}$.}
The $B_{\mathrm{h}}$ decrease \nna{is dominated} by the $B_{x}$ decrease, caused by the straightening of the inner legs of the pre-flare loops as shown by \citet{2012A&A...543A.110A}.
Then, $B_{\mathrm{h}}$ reaches the minimum and finally, $B_{\mathrm{h}}$ increases (e.g., for ROI-8 from $t=170t_{\mathrm{A}}$) until the end of the simulation. 
$B_{\mathrm{h}}$ increases \nna{after the reconnection took place and created the post-flare loops.} The post-flare loops are short and low-lying, therefore significantly more horizontal (lower inclination angle with respect to the solar surface) than the longer and higher loops formed later.
The $B_{\mathrm{h}}$ growth is due to the increase of the magnetic field parallel to the PIL (roughly $B_{y}$, Section~\ref{subsec:ch_bh}).
This is caused by the reconnection which transfers the different magnetic shear from the pre-flare loops to the post-flare loops \citep{2012A&A...543A.110A}.
 At the beginning of the simulation, the angle between the mean-PIL and the segment that joins the two foot-points, called the shear angle, is low (large shear).
During the flare, the shear angle slowly increases (the shear decreases).

%
The spatial distribution of $\delta B_{\mathrm{h}}$ shows a clear asymmetry with respect to the PIL. The additional concentration of the horizontal magnetic field occurs on the right side of the positive current ribbon, at around of $[x;y]\approx[0.5; 0.5]$ (Section~\ref{subsec:tem_evol_current}).
This effect is due to the CME deflection (see Figure~\ref{fig:f1}), which is a result of the system asymmetry \citep{2015ApJ...814..126Z}.
The CME is deflected toward the negative $x$-axis (Figure~\ref{fig:f1}), therefore $B_{x}$, \nna{initially negative in this area} increases at $[x;y]\approx[0.5; 0.5]$ (see Section~\ref{sec:mhd}).

Furthermore, the horizontal magnetic field decreases at the center of the magnetic polarities (Section~\ref{subsec:ch_bh}).
This result is consistent with observations \citep{2013SoPh..287..415P, 2017ApJ...839...67S}.

%
Moreover, the time difference between the beginning of the increase of $-j_{z}$ and $B_{\mathrm{h}}$ grows with time.
This is due to the fact that the shorter loops need less time to relax than the longer post-flare loops.

%
%
%
%

To compare the observation and the simulation we also focus on the time scales.
We suggest that our simulation covers a relative shorter physical time range than in the observational reports which present the magnetic field and the current variability several hours after the flare (e.g., \citet{2013SoPh..287..415P}).
To illustrate this we scale the dimensionless model to physical dimensions to estimate the duration of our simulation in the real time.
\nnd{We consider two possibilities, first, a young active region with a size of 50\,Mm (see e.g. AR11158 \citet{2011ApJ...738..167S}) and a coronal Alfv\'en speed of $c_{\mathrm{A}}=1000$ km s$^{-1}$, and then an old active region with a size of 200 Mm and $c_{\mathrm{A}}$=400 km s$^{-1}$.
%
The simulation shows a spatial scale of our active region of about 5 spatial units, the latter $L=1$ being defined as the distance between the PIL and the centre the one magnetic polarity at $z=0$.
Based on this information, we obtain the Alfv\'en time unit $t_{A}=L/c_{A}$=10\,s for a young active region (like AR11158), and 100\,s for an old spotless decaying active region.
These values suggest that the time between the start of the eruption ($t=165t_{\mathrm{A}}$) and the end of the simulation ($t=244t_{\mathrm{A}}$) is approximately 15\,min and 2\,h for a young and an old active region, respectively.
Our modeled duration of about 15\,min is consistent with the obseved duration of Bh increse of about 30\,min as reported for AR11158 (see Fig2, \citet{2017ApJ...839...67S}).}
%

%
\nna{Our analysis shows the increase of the photospheric horizontal magnetic field between the current ribbons.
The shape of the increasing horizontal magnetic field follows the expansion of the flare current ribbons.
Moreover, these two effects are confirmed by the recent high-cadence observation presented by \citet{2018arXiv181011733L}.
\mike{The analysis of our simulation and \citet{2018arXiv181011733L} observation suggest a strong spatial-relation between the evolution of the photospheric electric current and magnetic field during the flare.}
The most natural explanation of this process is the reconnection of magnetic field, but  the momentum conservation between the upward-moving CME and the underlying photosphere has also been proposed. 
If the momentum conservation was responsible for the flare process, the horizontal magnetic field increase would not be specifically located between the current ribbons, and its shape would not follow the expansion of the current ribbons.
Therefore, our model and the recent observation show that magnetic reconnection explains the dynamic of the horizontal magnetic field while the  CME momentum transfer hypothesis can not. 
}

In terms of magnetic field and current density evolution, our flare model is consistent with previous observations (Section~\ref{sec:intro}).
However, the current ribbon broadening was not noticed before.
It can be a numerical effect of the non-uniform mesh or physical due to increasing relaxation times for longer reconnected loops (as described above).

\section{\nna{Analysis of the Lorentz force densities}} \label{sec:lorentz_force_comp}

\subsection{The Lorentz force calculation} \label{sec:lorentz_calc}
Our simulation allows us to calculate the Lorentz force density (thereafter \nna{$\bm{f}$} and its components \nna{$f_x$, $f_y$, and $f_z$}) defined by

\begin{equation} \label{eq:jxb}
\bm{f}= \bm{j} \bm{\times} \bm{B}~.
\end{equation}

\nna{Alternatively, based on the vector magnetogram it is possible to estimate the total Lorentz force acting on the upper solar atmosphere.
In such case, the total Lorentz force is estimated as the surface integral coming from the volume integral of the Maxwell stress tensor \citep{2012SoPh..277...59F, 2012ApJ...759...50P}.
In our work, the total Lorentz force calculated with this alternative method (as presented above) is called the alternative Lorentz force ($\bm{F}$).
The vertical ($F_{z}$) and horizontal ($\bm{F}_\mathrm{h}$) components depend on both the vertical $(B_z)$ and horizontal ($\bm{B}_\mathrm{h}$) magnetic fields, that act on the bottom, upper and lateral surfaces (marked together as $A$) of the closed surface integral:}

\begin{equation} \label{eq:fvz_full}
F_{z}= \frac{1}{4 \pi} \oiint_{A} \frac{1}{2} \left( B_{z}^{2}-B_{\mathrm{h}}^{2}\right)\,dA~,
\end{equation} 

\begin{equation} \label{eq:fvh_full}
\bm{F}_\mathrm{h}= \frac{1}{4 \pi} \oiint_{A}\left(B_{z}\bm{B}_{\mathrm{h}}\right)\,dA~.
\end{equation}

\nnb{According to \citet{2012SoPh..277...59F}, we assume that the upper and lateral surfaces of the volume are significantly far from our active region, so the magnetic field contribution from these surfaces is negligible. 
Finally, only the bottom surface (the photosphere) will have a meaningful contribution because of the strong photospheric magnetic field.
This assumption allow us to estimate the total Lorentz force based only on the photospheric vector magnetic field, but only for physically isolated and flux-balanced photospheric domains (e.g. active regions).}

\nna{In our work, we focus on the sole integrand of $\bm{F}$ (thereafter the alternative Lorentz force density, $\bm{s}^\mathrm{ad}$). Its vertical ($s^\mathrm{ad}_{z}$) and horizontal components ($\bm{s}^\mathrm{ad}_\mathrm{h}$) are defined by:}
\begin{equation} \label{eq:fvz}
s^\mathrm{ad}_{z}=  \frac{1}{2} \left( B_{z}^{2}-B_{\mathrm{h}}^{2}\right)\,~,
\end{equation} 

\begin{equation} \label{eq:fvh}
\bm{s}^\mathrm{ad}_\mathrm{h}=B_{z}\bm{B}_\mathrm{h}\,~.
\end{equation} 
\nnb{Often, only $\bm{s}^\mathrm{ad}$ or $\bm{s}^\mathrm{ad}$ multiplied by a small surface element (e.g. area of one pixel) are used as the Lorentz force density to create Lorentz force density maps \citep{2012ApJ...759...50P}.}

\nna{The application of Lorentz force density maps as a diagnostic of the solar flares motivated us to test reliability of these maps.
%
%
Magnetic field observations allow us only to compute  $\bm{s}^\mathrm{ad}$ and not $\bm{f}$, while  numerical simulation gives us the magnetic field vector in a full 3D box, which \mike{allows} us to compute both $\bm{s}\mathrm{^{ad}}$ and $\bm{f}$.
Therefore, we first compare $\bm{s}^\mathrm{ad}$ obtained from the simulation and observations to determine whether the trends in $\bm{s}^\mathrm{ad}$ maps from our model are consistent with the trends in $\bm{s}^\mathrm{ad}$ maps obtained from observations.
 If they are, then we can use the simulation data to test how pertinent is the use of $\bm{s}^\mathrm{ad}$ instead of $\bm{f}$.}



\nna{First, we apply the method of \citet{2012ApJ...759...50P}, used in observational data analysis to calculate \nnb{$\bm{s}\mathrm{^{ad}}$ and $\bm{\delta s}^\mathrm{ad}$}.
We calculate \nnb{$\bm{s}^\mathrm{ad}_{z}$}, mesh point by mesh point over the entire field-of-view, considering the simulation time $\delta t$ of the analyzed frame and the onset of the eruption phase at  $t_{\mathrm{0}}=164t_{\mathrm{A}}$, and  we analyze $\bm{s}\mathrm{^{ad}}$ changes $(\delta s\,\mathrm{^{ad}})$ such as:}

\begin{figure*}[ht!]
\epsscale{1.2}
\plotone{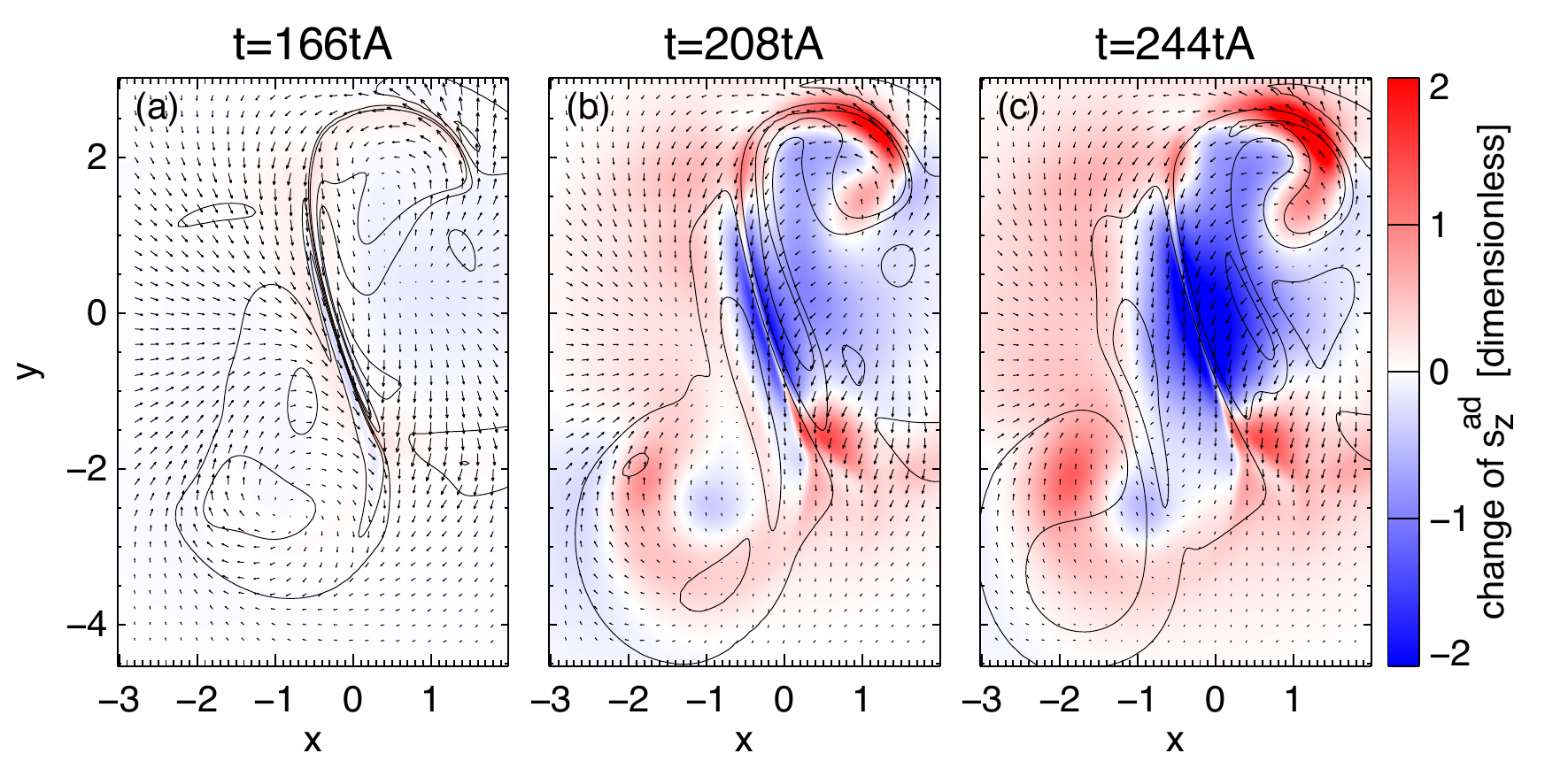}
\caption{
The vertical component of \nna{the alternative Lorentz force density} changes (\nna{$\delta s^\mathrm{ad}_{z}$}) at three different times during the eruption phase. 
The \nnz{change of} \nnb{$s^\mathrm{ad}_{z}$} is defined by the color-code. Arrows and electric current density contours are the same as in Figure~\ref{fig:f3}. 
\\
(An \href{run:./md.mp4}{animation}  of this figure is available in the online journal.)
\label{fig:f8}}
\end{figure*}
\begin{equation} \label{eq:chfr}
\delta F_{z}=\frac{1}{4 \pi} \oiint_{A} \underbrace{ \frac{1}{2}\left( \delta B_z^{2}- \delta B_\mathrm{h}^{2}\right)}_{\delta s_{z}^\mathrm{ad}}\,dA~,
\end{equation} 
for the fixed height above the photosphere, where:

\begin{equation} \label{eq:chbh}
\delta B_{i}^{2}=B_{i}^{2}(\delta t)-B_{i}^{2}(t_{0})\mathrm{~with~}i=h;z.
\end{equation}

\nna{In Figure~\ref{fig:f8}, we present maps of the temporal evolution of $\delta s^\mathrm{ad}_{z}$ at the photosphere ($z=0.1$).
In the region located between the straight part of the current ribbons, we notice a significant increase of the negative $s^\mathrm{ad}_{z}$.
Additionally, a slight increase of the negative \nnb{$s^\mathrm{ad}_{z}$} also exists at $[x;y]\approx[-1; -2.5]$.
In general, the positive \nnb{$s^\mathrm{ad}_{z}$} slightly increases in the rest of the map, except in the northern region near the hook ($[x;y]\approx[1; 2.5]$), where the increase of the positive \nnb{$s^\mathrm{ad}_{z}$} is significant.} 

\nnb{Previous observational analysis showed a significant increase of the negative \nnb{$s^\mathrm{ad}_{z}$} around the PIL in the photosphere (see Section~\ref{sec:intro}). 
This implies that \nnb{$s^\mathrm{ad}_{z}$} in our model is consistent with \nnb{$s^\mathrm{ad}_{z}$} obtained from previous observations.
Therefore, we can compare $\bm{s}^\mathrm{ad}$ and $\bm{f}$ obtained from our model.}

\subsection{\nna{Comparison of the alternative Lorentz force density with the Lorentz force density and the horizontal magnetic field}} \label{sec:lorentz_comp_rois}

We compare the temporal variability of \nna{$\bm{f}$} and \nnb{$\bm{s}\mathrm{^{ad}}$} for the ROIs swept by the current ribbon of the negative polarity (Figure~\ref{fig:f2}).
For this purpose, we calculate the average values of \nna{$\bm{f}$} and \nnb{$\bm{s}\mathrm{^{ad}}$}, at each time \nna{of the simulation}, for each ROI as a weighted arithmetic mean, where the weights are the mesh pixel area.

\begin{figure*}[ht!]
\epsscale{1.2}
\plotone{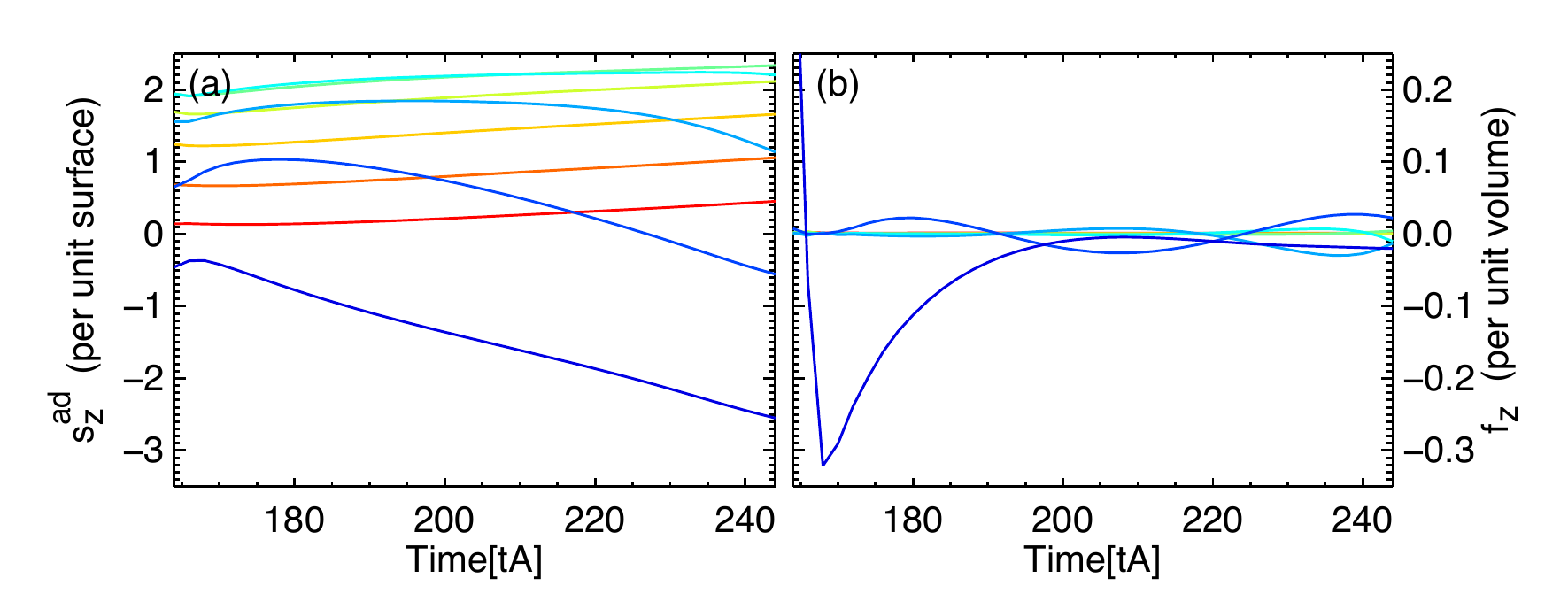}
\caption{Comparison of temporal variability of \nna{the alternative Lorentz force density} \nnb{$s^\mathrm{ad}_{z}$} (a) and the Lorentz force density \nna{$f_{z}$} (b) at $z=0.1$. The color-coded lines indicate the regions-of-interest (boxes 0-9) in Figure~\ref{fig:f2} in which these quantities are calculated. \nnz{All units are dimensionless.}\label{fig:f6}}
\end{figure*}

Figures~\ref{fig:f6}a and b present clear differences between \nna{$f_z$} and \nnb{$s^\mathrm{ad}_{z}$} in terms of the temporal \nna{evolution} and the magnitude.
These trends are clearly visible for all ROIs.
We analyze separately the ROIs which are swept by the current ribbon (ROI $\geq 5$) and which are not (ROI $< 4$).

For ROIs 0-4, \nna{$f_z$} stays almost constant with time (Figure~\ref{fig:f6}b) while \nnb{$s^\mathrm{ad}_{z}$} continuously increases (Figure~\ref{fig:f6}a).
For ROIs 5-8, \nna{$f_z$} (Figure~\ref{fig:f6}b) shows clear oscillations which are observed during the whole simulation time. 
The oscillation amplitude is higher for the ROIs closer to the PIL.
The discontinuity observed in the \nna{$f_z$} curve before $t=170t_{\mathrm{A}}$ is the result of the ROI-8 location on the PIL.
For ROIs 5-8, \nnb{$s^\mathrm{ad}_{z}$} (Figure~\ref{fig:f6}a) slightly increases before these ROIs are swept by the current ribbon (see, e.g., \nnb{$s^\mathrm{ad}_{z}$} for ROI-7 [blueish curves], before $t=175t_{\mathrm{A}}$).
When the ROI is swept by the current ribbon, \nnb{$s^\mathrm{ad}_{z}$} reaches its maximum (e.g., for ROI-7, at $t=180t_{\mathrm{A}}$).
After the ROIs have been swept by the current ribbon, \nnb{$s^\mathrm{ad}_{z}$} clearly decreases and this trend continues until the end of the simulation.
%

\begin{figure*}[ht!]
\epsscale{1.2}
\plotone{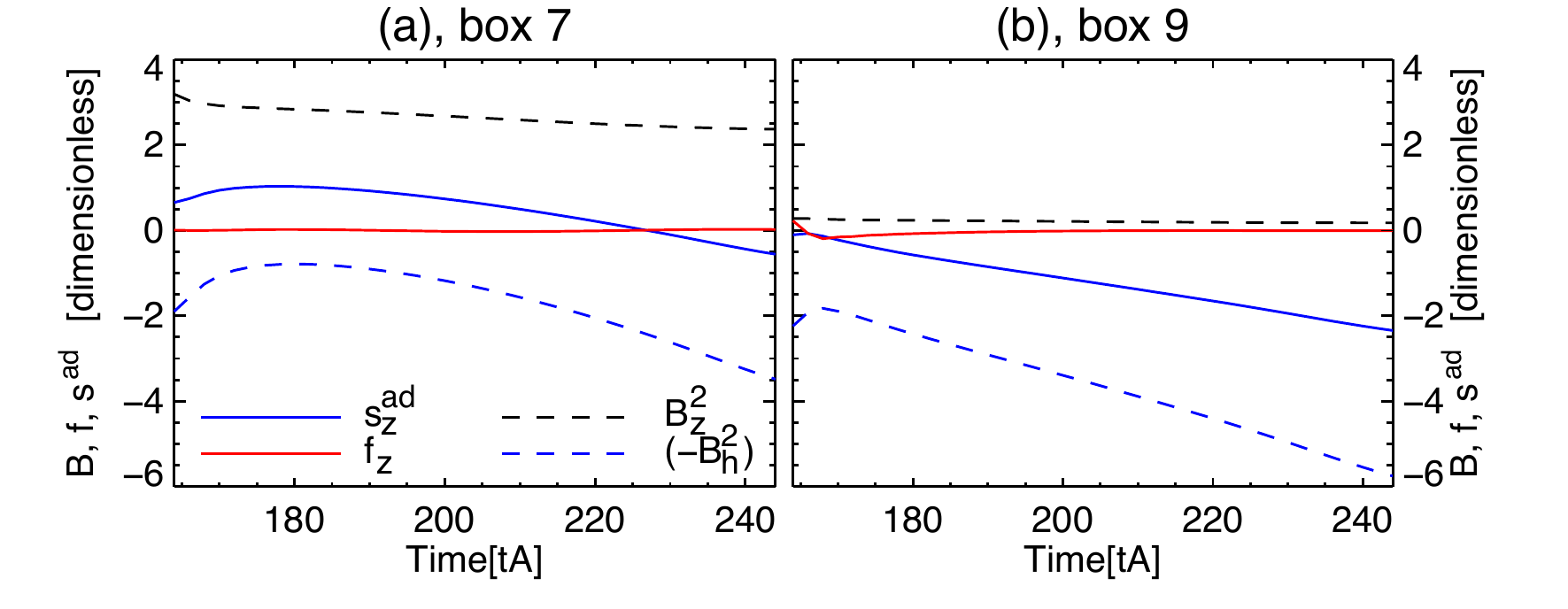}
\caption{Temporal variability of \nna{the alternative Lorentz force density $s^\mathrm{ad}_{z}$} and the individual terms \nnb{($\bm{B}^{2}_{h}, B^{2}_{z}$)} used to calculate it. For comparison, the diagram shows also \mike{$f_{z}$ ($|\bm{j}\bm{\times}\bm{B}|_{z}$)}. These quantities are provided for the same boxes 7 (panel a) and 9 (panel b) as presented in Figure~\ref{fig:f2}.  \label{fig:f7}}
\end{figure*}

\nna{Equation~\ref{eq:fvz} shows that the vertical component of the alternative Lorentz force density ($s^\mathrm{ad}_{z}$) depends on $B_{z}$ and $B_{\mathrm{h}}$.
To determine which term ($B_{z}$ or $B_{\mathrm{h}}$) dominates in  Equation~\ref{eq:fvz}, we compare the temporal evolution of $B_{z}^2$, $B_{\mathrm{h}}^2$ and $s^\mathrm{ad}_{z}$ in two ROIs.
We choose the ROI located in different domains of the flare to avoid the local trends and to build a more general view of the relation between \nnb{$s^\mathrm{ad}_{z}$} and the magnetic field.}
\nna{We analyze the region which is swept by the current ribbon -- ROI-7 (Figure~\ref{fig:f7}a) -- and the region located at the PIL -- ROI-9 (Figure~\ref{fig:f7}b).
In both ROIs, the $B_{\mathrm{h}}^2$ curves are almost parallel to \nnb{$s^\mathrm{ad}_{z}$}, while the $B_{z}^2$ curves stay almost constant.
This shows that the temporal evolution of $s^\mathrm{ad}_{z}$ is dominated by $B_{\mathrm{h}}$. 
}

\nna{In addition, the maps of the $\delta s^\mathrm{ad}_{z}$ (as presented in Section~\ref{sec:lorentz_calc}) show a striking resemblance to the $\delta B_{\mathrm{h}}$ maps (Figure~\ref{fig:f3}). 
However, their signs are opposite. This suggests that $s^\mathrm{ad}_{z}$ is just another way to see $B_{\mathrm{h}}$.
Retrospectively, this is to be expected from Equation~\ref{eq:chfr} as well as from the analysis of Figure~\ref{fig:f7} which shows that \nnb{$s^\mathrm{ad}_{z}$} mainly depends on $B_{\mathrm{h}}$.}

\subsection{The Lorentz force density in the corona} \label{sec:lorentz_perpentic}
In observational data analysis, the Lorentz force density can be calculated only in the photosphere, because of the limitation of the vector magnetic field measurements.
\nnb{The coronal magnetic field is much weaker than in the photosphere.
Additionally, the spectral lines in the corona are dimmer and broader comparing to the photospheric spectral lines. These limitations make it impossible to analyze the $s^\mathrm{ad}_{z}$ distribution in the solar corona.
However, our simulation is free from these limitations.}
To understand the solar flare evolution, it is crucial to investigate the Lorentz force changes also in the upper solar atmosphere. 
3D MHD simulations give us this opportunity.
\nna{Additionally, it allows us to evaluate how important is the influence of the lateral and top boundaries of the system to the calculation of $\bm{s}\mathrm{^{ad}}$.}

\begin{figure*}[ht!]
\epsscale{1.2}
\plotone{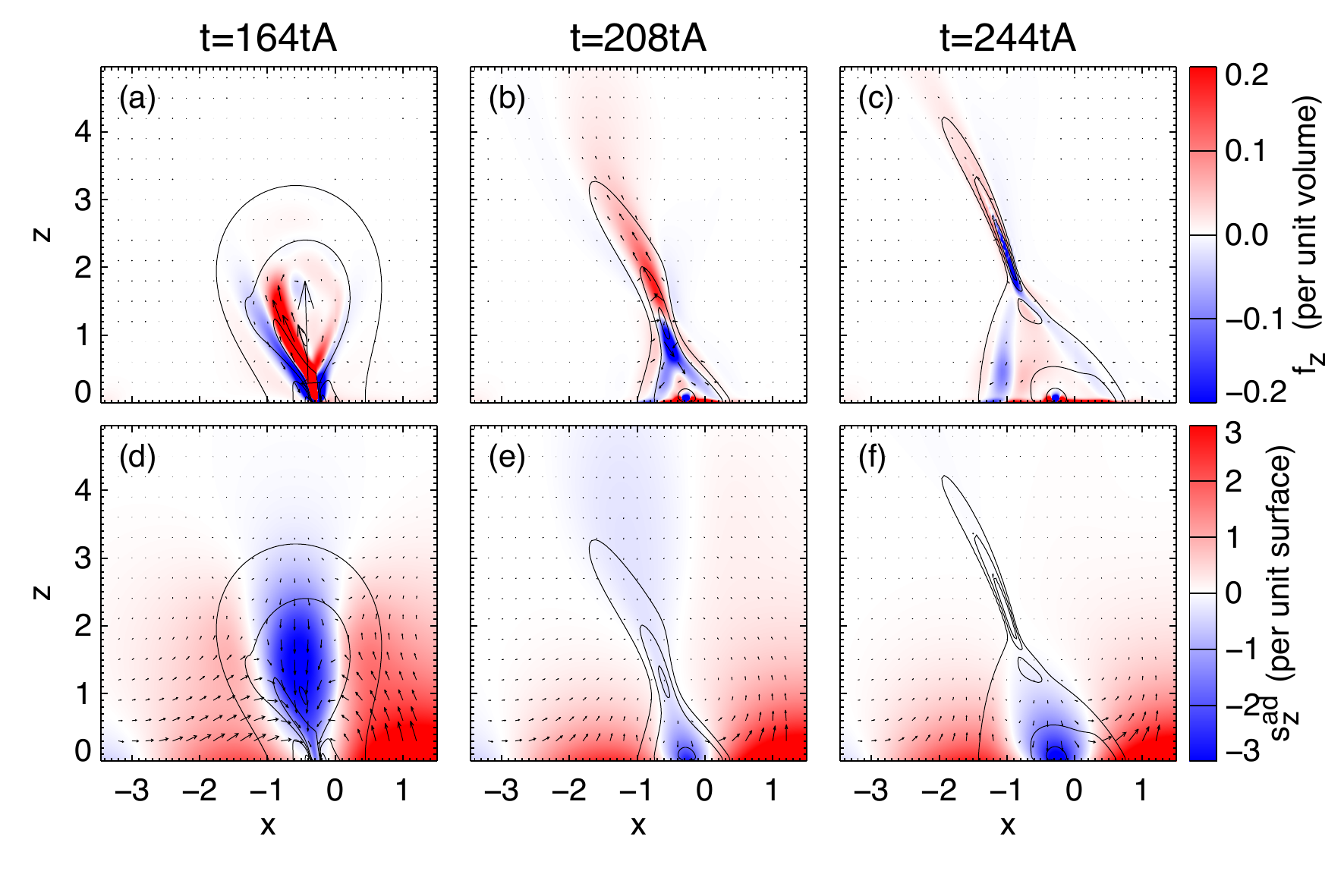}
\caption{Comparison of the Lorentz force density \nna{$f_{z}$} (a-c) and \nna{the alternative Lorentz force density} \nnb{$s^\mathrm{ad}_{z}$} (d-f) at three different times during eruption phase, view along a vertical cut at $y=0$. The vertical component is presented in the color-scale.  Vectors indicate \nna{$f_{x}$, $f_{z}$} (panels a-c) and \mike{$s^\mathrm{ad}_{x}$}, \nnb{$s^\mathrm{ad}_{z}$} (panels d-f). Vectors \nna{$f_{x}$, $f_{z}$} are multiplied by a factor of 5. The black contour lines mark current density at levels $|\bm{j}|$=1.1, 3.0, 6.0 (see Figure~\ref{fig:f10})  and \mike{show an} inverted Y-shape. \nnz{All units are dimensionless}.
\\
(An \href{run:./me.mp4}{animation}  of this figure is available in the online journal.)
\label{fig:f9}}
\end{figure*}

Based on the method described in Section~\ref{sec:lorentz_calc}, we calculate \nna{$f_{z}$} and \nna{$s^\mathrm{ad}_{z}$} \mike{in} the vertical ($x$, $z$) plane at $y=0$, which is located at the surface that separates the negative and positive  polarities.
\nna{To calculate $s^\mathrm{ad}_{z}$, we assume that the solar corona is built of layers parallel to the solar photosphere. At each height ($z$), the meaningful contribution to $s^\mathrm{ad}_{z}$ comes from the magnetic field of the layer at that height, the influence of the upper and bottom layers being negligible.}
The comparison of $f_{z}$ and \nnb{$s^\mathrm{ad}_{z}$} is presented in Figure~\ref{fig:f9}.

%
Before the flare eruption (Figure~\ref{fig:f9}a), the upwards $f_{z}$ distribution has a tear-drop shape structure ($x\in[-1.1; 0.1]$ and $z\in[0; 2.1]$) with the almost hollow-core ($[x;z]\approx[-0.3; 1.7]$).
This hollow-core corresponds to the flux rope position (Figure~\ref{fig:f1}a).
The whole structure is co-spatial and has a high correlation with the electric current density (see black contours). 
This link between the Lorentz force density and current sheets was already noted in \citet{2005A&A...444..961A} for a non-erupting simulation of the current sheet formation in the quasi-separatrix layers.
The tear-drop shape structure \mike{($[x;z]\approx[-0.3; 0]$) is rooted} at the solar surface (boundary layer) between the current ribbons.
\nna{The Lorentz force density of the Y-shape structure at the bottom of the tear-drop structure is positive in the inner part and is bounded by negative values.} 
%
These downward force distributions are rooted at both current ribbons at the solar surface ($[x;z]\approx[-0.4; 0]$ and $[x;z]\approx[-0.2; 0]$).
In the rest of the map, the vertical Lorentz force is negligible.
In our simulation, the boundary layer (solar surface) is numerically defined \citep{2015ApJ...814..126Z} and we do not use any observational data to calculate it.
\nna{Figure~\ref{fig:f9}a-c shows that above $z\gtrsim0.1$, $f_z$ varies smoothly, while boundary effects are visible for smaller $z$.}
This indicates that the choice $z=0.1$ for earlier maps is suitable.


At the HFT (the 3D version of the X point) the flare reconnection starts at $t=165t_{\mathrm{A}}$ ($[x; y]\approx[-0.9; 1.2]$ in Figure~\ref{fig:f9}b).
It creates new highly-curved field lines above and below the HFT. 
These curvatures induce magnetic tension, hence Lorentz forces, and flare reconnection jets.
Figure~\ref{fig:f9}b clearly shows the upward \nna{$f_z$} jet above the HFT (blue color concentration at $[x;z]\approx[-0.4; 1]$) and the downward Lorentz forces below the HFT (red color concentration at $[x; z]\approx[-0.7; 2]$) as well as inflows (see arrow) on the left ($x<-0.6$ and $y\approx1.3$) and on the right ($x>-0.6$ and $z\approx1.3$) side of the HFT.
As reconnection proceeds, more and more flare loops are formed so the reconnection happens at higher and higher altitudes, hence the rise of the HFT and the dual \mike{up-downward $f_z$} (Figure~\ref{fig:f9}c).
That is consistent with the standard CSHKP model \nna{extended to 3D \citep{2010ApJ...708..314A, 2013A&A...555A..77J}.}
In the rest of the map, $f_{z}$ is still negligible.

\nna{The alternative Lorentz force density, $s^\mathrm{ad}_{z}$}, presents a significantly simpler pattern than \nna{$f_{z}$}.
Before the start of the eruptive phase (Figure~\ref{fig:f9}d), the inner part of the flare (contour at $|\bm{j}|$=1.1) characterizes a strongly downward \nnb{$s^\mathrm{ad}_{z}$} ($[x;y]\approx[-0.3; 1.5]$) also within the flux rope.
There is also a lack of a reconnection-jet related to \nnb{$s^\mathrm{ad}_{z}$} during the eruption.
However, the rest of the map presents an upward \nnb{$s^\mathrm{ad}_{z}$}.
During the flare, the distribution of the downwards \nnb{$s^\mathrm{ad}_{z}$} concentration moves toward the photosphere and the magnitude \nnb{$s^\mathrm{ad}_{z}$} increases (Figure~\ref{fig:f9}e, f) there.
This trend continues until the simulation ends.

The comparison of \nna{$f_z$} (Figure~\ref{fig:f9}a-c) with \nnb{$s^\mathrm{ad}_{z}$} (Figure~\ref{fig:f9}d-f) shows a lack of similarity between them, from the start of the eruption until the end of the simulation.
 \nna{$f_z$} is completely different from \nnb{$s^\mathrm{ad}_{z}$}.
 \nnb{$f_z$ is weak only at the cusp edge while \nnb{$s^\mathrm{ad}_{z}$} is strong in the whole domain; $s^\mathrm{ad}_{z}$ is strongly negative around of the current sheet and positive outside.}
 The \nnb{$s^\mathrm{ad}_{z}$} distribution characterizes a lack of the reconnection-jet despite the reconnection in the model.
 %

\subsection{Discussion of the Lorentz force densities} \label{sec:lorentz_discussion}
\nna{The alternative Lorentz force density ($\bm{s}^\mathrm{ad}$)} is used in the observational data analysis of the solar photosphere as \mike{an} approximation \mike{to} the Lorentz force density.
Based on the simulation, we find that the downward \nnb{$s^\mathrm{ad}_{z}$} significantly increases in the region between the current ribbons, during the flare (\nnb{Section~\ref{sec:lorentz_calc}} and Section~\ref{sec:lorentz_comp_rois}).
The \nnb{$s^\mathrm{ad}_{z}$} changes are closely related to the changes of the horizontal magnetic field (Section~\ref{sec:lorentz_comp_rois}).
 The previous observational analyzes show a significant increase of negative \nnb{$s^\mathrm{ad}_{z}$} around the PIL in the photosphere (Section~\ref{sec:intro}). 
This implies that \nnb{$s^\mathrm{ad}_{z}$} computed from our model is consistent with \nnb{$s^\mathrm{ad}_{z}$} obtained from previous observations.
The measurement of the magnetic field in the solar corona is highly limited because the coronal magnetic field is much weaker than in the photosphere.
Additionally, the spectral lines in the corona are dimmer and broader comparing to the photospheric spectral lines.
These limitations make it impossible to analyze the \nnb{$s^\mathrm{ad}_{z}$} distribution in the coronal observations.
However, our simulation is free from these limitations.

We use our simulation to study \nna{$\bm{f}$} and \nnb{$\bm{s}^\mathrm{ad}$} in the photosphere (Section \ref{sec:lorentz_comp_rois}) and in the solar corona (Section~\ref{sec:lorentz_perpentic}).
Immediately before the reconnection, the flux rope generates the upward \nna{$f_z$} distribution of the tear-drop shape \nna{(red color in Figure~\ref{fig:f9})}. 
The evolution of the $\Omega$-shaped arcade creates the downward \nna{$f_z$} distribution outside of the bottom border of the tear-drop shape structure (blue color on Figure~\ref{fig:f9}).
Then, the reconnection begins at the HFT and the flux rope expands.
\nna{The reconnected magnetic field lines are characterized by a strong magnetic tension in the exhaust  which drives the reconnection jet above and below the HFT (Figure~\ref{fig:f9}b at $z>1.2$).
This triggers a downward and an upward flows moving away from HFT (Figure~\ref{fig:f10}h, i).}
The magnetic field pressure decreases around the HFT and allows more magnetic field to go into the current sheet.
The curvature of the field lines around the HFT generates a Lorentz force on the right and left side of the HFT (Figure~\ref{fig:f9}b at $x\in [-0.9; -0.1]$ and $z\approx1.2$).
This Lorentz force causes the inflow (arrows in Figure~\ref{fig:f10}g-i at $x\in [-0.9 ;-0.1]$ and $z\approx1.2$ ) of the coronal magnetic field into the current sheet \citep{2017ApJ...837..115Z}.
The tension forces make the post-flare loops to \nna{shrink} down, and at some point they relax and decelerate.
This deceleration comes from the upward \nna{$f_z$}, which is due to the line-tying (i.e., the impenetrable photosphere/the line-tied boundary at $z$=0).
This complex behavior of the upwards and downwards \nna{$f_z$} around the cusp and the cusp expansion along the $x$-axis are responsible for the damped oscillation pattern of the photospheric \nna{$f_z$} in \mike{ROIs} 5-8 (Section~\ref{sec:lorentz_comp_rois}).

The comparison of \nnb{$\bm{f}$} and \nnb{$\bm{s}^\mathrm{ad}$} points to strong differences between them, both in the photosphere (Section~\ref{sec:lorentz_comp_rois}) and in the corona (Section~\ref{sec:lorentz_perpentic}).
The maps of \nna{$f_z$} and \nnb{$s^\mathrm{ad}_{z}$} in the corona show completely different structures.
%
%
Additionally, \nnb{$s^\mathrm{ad}_{z}$} is directly related to $B_{\mathrm{h}}$ while \nnb{$f_z$} is not (Section~\ref{sec:lorentz_comp_rois}). 
Thus, these discrepancies imply that the \nna{alternative Lorentz force density, $\bm{s}\mathrm{^{ad}}$}, is not a good approximation of the Lorentz force density, \nna{$f_z$}.
\nna{Mathematically, the Gauss-Green-Ostrogradski theorem transforms the (coronal) volume integral of the divergence of the Maxwell stress tensor into a closed surface integral. Assuming that the upper and lateral surfaces of the integral are far enough from the investigated active region, their magnetic field contribution is negligible and the only significant contribution  comes from the lower surface (photosphere). However, even under these conditions, the integral over the (photospheric) surface is meaningful whereas the spatial distribution of its integrand, i.e., the alternative Lorentz force density, is a priori not a rigorous proxy for the Lorentz force density. Indeed, the map of the  alternative Lorentz force density presents each element (pixel or mesh-point) as an individual domain, which clearly breaks the above assumption. Moreover, the multiplication of each element of the map of the alternative Lorentz force by the size of that element does not solve a problem. Still, the assumption is not fulfilled, even though the map presents a correct unit of force. Only the integration of the magnetic field contribution from all individual elements at the whole photospheric surface of the investigated domain fullfils the above assumption. }

\nna{Thus, in a closed surface integral (Equations~\ref{eq:fvz_full},  \ref{eq:fvh_full}) only the integral calculated under the conditions defined by \citet{2012SoPh..277...59F} is meaningful and presents the \mike{total} Lorentz force, not the integrand.
This is fully consistent with our empirical test, showing that the alternative Lorentz force density, $\bm{s}\mathrm{^{ad}}$, is not a good approximation of the Lorentz force density, $\bm{f}$}. 
Therefore, previously published results of \nnb{$s^\mathrm{ad}_{z}$ and especially the so-called Lorentz force maps ($s^\mathrm{ad}_{z}$, $\delta s^\mathrm{ad}_{z}$ maps)} must be taken with caution.

\section{Coronal reconnection driven by the photospheric field} \label{sec:induction_equation}
\subsection{\nna{Magnetic induction with line-tied flare-flows}} \label{sec:spe_parallel}
\begin{figure*}[ht!]
\epsscale{1.05}
\plotone{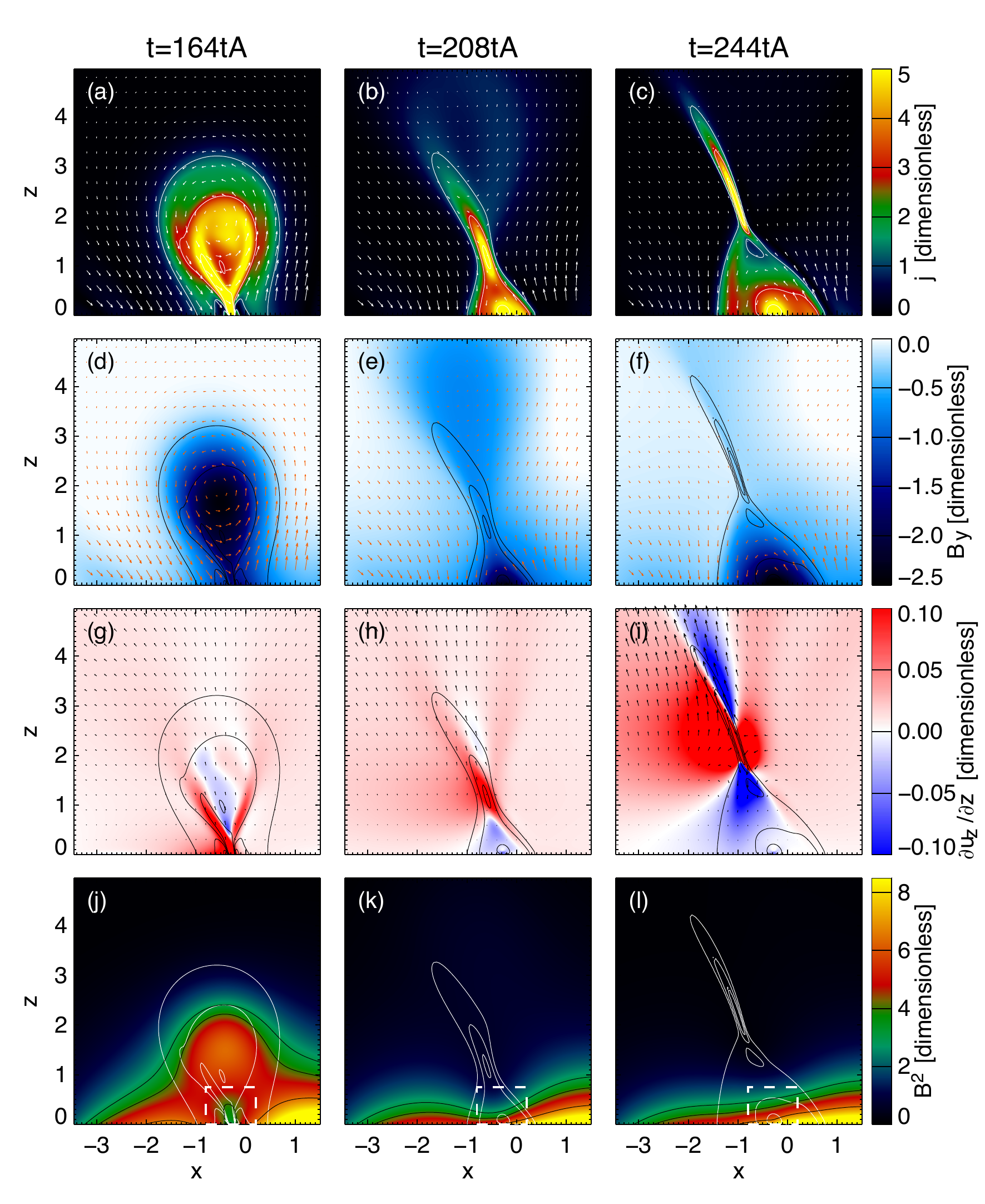}
\caption{The evolution of the eruptive event in different observables, viewed in the ($x$, $z$) plane at $y=0$. The temporal evolution of the magnitude of the current density (a-c) is presented by the color scale. \nnc{Contours (white: a-c, j-l; black: d-i) are at threshold $|\bm{j}|$=1.1, 3.0, 6.0, and identify the inverted Y-shape}. Arrows mark the $B_{x}$ and $B_{z}$ components of the magnetic field (a-f). The temporal evolution of $B_{y}$ (d-f), defined by the blue/white color-scale, is almost unipolar. The velocity derivative $\partial{u_{z}}/\partial{z}$ (g-i) shows a clearly change of sign in the inverted Y-shape structure. The velocity components $u_{x}$, $u_{z}$ are indicated by arrows (g-i). \nnc{The temporal evolution of the magnetic field energy density (j-l) is presented by the color scale. Contours (black: j-l) are at threshold $B^2$=3.0, 4.0, 6.5. The dashed line box (j-l) marked the region of the increase magnetic energy density within the volume of cusp that contains the flare loops (compare k and l). }
\\
(An \href{run:./mf.mp4}{animation}  of this figure is available in the online journal.)
\label{fig:f10}}
\end{figure*}

The simulation allows us to study the properties of flare evolution from the photosphere to the corona.
In  Section~\ref{subsec:b_para}, we show a clear increase of $B_{y}$ near the PIL, between the current ribbons.
From morphological arguments, we suggest that this increase is related to the change of the field line configuration as the \nna{reconnected} field lines become the \nna{post-flare} loops (Section~\ref{subsec:b_para}).
Here, \nna{we use the first-principle physical ingredients from ideal MHD, the ideal induction equation:}
\begin{equation} \label{eq:ind_eq}
\frac{\partial \bm{B}}{\partial t} = \bm{\nabla} \bm{\times} \left( \bm{u} \bm{\times} \bm{B} \right),
\end{equation}
\nna{to explain the $B_y$ increase.}
\nna{In the vertical plane ($x$, $z$) at $y=0$ of the solar atmosphere, the} flow analysis shows that the plasma velocity ($\bm{u}$) is very small (\nna{arrows} in Figure~\ref{fig:f10}g) immediately before the eruption.
The line-tying implies that for an idealised photospheric plane, $u_z$=0 and $u_{\perp}$ is not influenced by the coronal evolution.
 Photospheric motion with finite $u_{\perp}$ can only be driven by slow local drivers (such as pressure gradients) or by equally slow subsurface drivers (such as convection and flux emergence).
On fast eruption time-scales, these drivers can be neglected, so $u_{\perp}=0$ just like $u_{z}$, as prescribed in the simulation.
In our model, the line-tied plane is at $z=0$, but in this paper, we analyze the magnetic fields and electric currents at $z=0.1$, which is twenty times lower than the altitude of the axis of the pre-eruptive flux rope. 
At this small altitude, the effect of the line-tied boundary is still very strong, so the plasma velocities remain very small in the simulation.
Therefore, we can assume for simplicity that $\bm{u}(z=0.1)=0$. 
While this implies that $\bm{\nabla}_{\perp}\bm{u_i}$ is also zero, it does not prevent the vertical gradient  ($\partial/\partial{z}$) from having finite values. 
After developing Equations~\ref{eq:ind_eq} and cancelling out the null terms we obtain:

\begin{equation} \label{eq:ind_eq_bz}
\frac{\partial B_z}{\partial t} = 0 \implies B_{z}=\text{const},
\end{equation} 

\begin{equation} \label{eq:ind_eq_by}
\frac{\partial B_{\perp}}{\partial t} =- B_{\perp} \frac{\partial u_z}{\partial z}\mathrm{~where}\perp = x,y.
\end{equation} 
Based on Equations~\ref{eq:ind_eq_bz} - \ref{eq:ind_eq_by}, we study the temporal evolution of the variables.
Figure~\ref{fig:f10} shows the evolution of the current density ($|\bm{j}|$), the $y$-component of the horizontal magnetic field ($B_{y}$) and the spatial partial derivative of the flow  ($\partial{u_z}/\partial{z}$) in the vertical ($x$, $z$) plane.

Immediately before the eruption (Figure~\ref{fig:f10}a), we notice the characteristic ``filled $\Omega$-shaped'' structure of a high current density concentration that surrounds the pre-eruptive flux-rope ($[x; z]\approx[-0.3; 2]$). %
Inside this structure, the current density is even higher and has a tear-drop shape structure.
During the eruption (Figure~\ref{fig:f10}b), the usual cusp shape has been formed at around $t=208t_{\mathrm{A}}$ with the clearly visible flare current sheet ($[x; z]\approx[-0.5; 1.5]$ at $t=208t_{\mathrm{A}}$).
\nna{Later, the cusp expands upwards (Figure~\ref{fig:f10}c), \mike{and} its footpoints are co-spatial with the current ribbons (Figures~\ref{fig:f1} and  \ref{fig:f3})}.

Initially, a high concentration of negative $B_{y}$ ($x\in[-1; 0.4]$ and $z\in[0; 3]$ ) \nna{ is observed } in the region between the current ribbons and nearby the flux rope (Figure~\ref{fig:f10}d).
The $B_{y}$ concentration nearby the current ribbons ($[x; z]\approx[0.3; 0.1]$) continuously increases during the eruption (Figure~\ref{fig:f10}e, f).
Before the eruption, the $B_{y}$ concentration around the flux rope ($[x; z]\approx[-0.3; 2]$) forms the tear-drop shape structure, then moves upwards and at the same time decreases (Figure~\ref{fig:f10}d).
Finally, this $B_{y}$ concentration moves out of our field-of-view (Figure~\ref{fig:f10}e).
In general, the strong increase of $B_{\mathrm{h}}$ (and $B_y$) and therefore the magnetic field energy density (although we do not show it in this paper) exists in the whole cusp, hence all along the flare loops.
On the contrary, the $B_z$ distribution varies significantly less in the cusp.

Before the eruption (Figure~\ref{fig:f10}g), the $\partial{u_{z}}/\partial{z}$ map shows a ``Y-shaped structure" of $\partial{u_z}/\partial{z}>0$, below the flux-rope ($x\approx-0.3$ and $z<2$) related to the current density concentration (\nna{black or white contours, in Figure~\ref{fig:f10}}).
When the eruption starts the cusp ($[x; z]\approx[-0.5; 0.5]$) with $\partial{u_z}/\partial {z}<0$ appears below the ``Y-shape structure'' (Figure~\ref{fig:f10}h).
This cusp is roughly co-spatial with the current density concentration (Figure~\ref{fig:f10}b).
It expands further and creates a double Y-shape structure (Figure~\ref{fig:f10}i).
On the left and on the right side of the current sheet (black contours, in Figure~\ref{fig:f10}), $\partial{u_z}/\partial{z}$ \nna{is positive}, but above and below the current sheet $\partial{u_z}/\partial{z}$ \nna{is negative}. 
The expansion of the double Y-shape continues until the end of the simulation (Figure~\ref{fig:f10}i).

\subsection{\nna{Amplification of horizontal fields}} \label{sec:spe_analize}

%
\nna{Once the flare has started, magnetic field reconnects at the HFT, leading to the formation of post-flare loops.}
The pile-up of the  magnetic field lines below the HFT  is the result of a slow coronal relaxation of the reconnected flare loops, which initially go down due to the magnetic tension at their apex, and which gradually find a force-free equilibrium all along their length, from the corona down to the photosphere. 
In this process, the forces come from the local reconnection region, not from a reaction to the bulk/extended eruption of the CME.
%

The $B_{y}$ component of the horizontal magnetic field increases not only at the surface around the PIL (Section~\ref{sec:mf_cur_var}) but also in the whole interior of the cusp (Section~\ref{sec:spe_parallel}).
This is due to the post-flare loop orientation, which is almost parallel to the $y$-axis, especially the post-flare loops created at the beginning of the eruptive phase.

\mike{The relaxation of the flare loops and braking as they arrive at the photosphere means that the local values of $\partial{u_z}/\partial{z}$ are negative (Figure~\ref{fig:f10}g; Section~\ref{sec:spe_parallel}).}
 The values of $u_z$ decrease with $z$, so a negative $u_z$ becomes more and more negative and finally $u_z \approx 0$ at $z \approx 0$.

Here, we focus on the induction equation (Equation~\ref{eq:ind_eq_by}).
Our discussion is provided for $\perp = y$, but the analysis for $\perp = x$ can be conducted in the same fashion. 
Figures~\ref{fig:f10}d-f show $B_y<0$ in our full-field-of-view.
If $B_{y}$ and $u_z$ are negative and $u_{z}$ is stopped at low $z$ then $\partial{u_z}/{\partial z}$ is also negative (the post-flare loops shrink).
Under these circumstances, Equation~\ref{eq:ind_eq_by} implies $\partial{B_y}/{\partial t}<0$. 
Moreover, if $\partial{B_y}/{\partial t}<0$ and $B_y<0$ then $|B_{y}|$ increases. 
This result can be generalised also for $|B_x|$. 
Finally, we conclude that the increase of the horizontal magnetic field around the PIL has a fundamental explanation based on the induction equation.
\nnc{Moreover, the volume decrease is not causing or caused by the magnetic energy decrease. In this region, the ideal contraction of the flare loops leads to a diminishing volume while $B_{y}$ (and also $B_{x}$) locally increases through flux conservation (see Figure~\ref{fig:f10}d-f in XZ). This magnetic field increases lead to a local increase of the magnetic energy density (see dashed line box in Figure~\ref{fig:f10}k-l). This process naturally exists in contracting parallel field lines. It shows how the volume decrease of flare loops is associated with local energy increase. This local increase does not contradict the total magnetic energy decrease through reconnection and the CME expansion (see Figure~\ref{fig:f10}k-l as well as the Figure 9 in \citealt{2015ApJ...814..126Z}). There is a local decrease of the magnetic field energy density above the PIL in a first stage (see dashed line box in Figure~\ref{fig:f10}j-k). It is due to the flux rope and it is surrounding overlying sheared loops being taken away in the CME. Then on a second stage (see dashed line box in Figure~\ref{fig:f10}k-l) an increase occurs where the flare loops form at the former position of the previous eruptive flux rope. While the evolving full distributions of $B^2$ are complex because they involve all three magnetic field components and all field lines surrounding the eruption, the two aforementioned staged are related with the evolution of the shear component $B_y$, firstly its expulsion within the flux rope, then its amplification  within the contracting flare loops.}


\section{Summary and conclusion}
\label{sec:sumcon}

In this paper, we address the physical origins of the long-duration increasing horizontal magnetic-fields and downward Lorentz-force \mike{values of} the Sun's photosphere around the polarity-inversion-line (PIL) as measured with vector magnetograms during solar eruptions. To do so, we analyzed a generic zero-$\beta$ MHD simulation of eruptive flares that was not designed to model these behaviors a priori. 

We used the same methodology as in observational and data analyses, and recovered the general observed properties. We also related the synthetic observables with various quantities that are not available or difficult to extract from observations, but that could be extracted from the three-dimensional simulation. In particular, we considered the dynamics of the area within which horizontal magnetic fields increase, its link with the coronal dynamics, and with the development of electric current density concentrations. Also for the first time, we compared \mike{$\bm{j}\bm{\times}\bm{B}$} with \nna{the alternative Lorentz force density used in observations.}

Our analyses of the spatio-temporal evolution of magnetic fields, electric currents, forces and flows led us to question previously published interpretations for photospheric horizontal magnetic-field increases during eruptions.

\mike{The region where the horizontal magnetic field increases in the photosphere in the simulation is mainly around the PIL. This matches observations.} These regions gradually expand between electric-current ribbons, while they spread away from the PIL. Since the latter are an MHD signature of flare ribbons, this implies a link with the flare rather than with the CME. 

The increases of $B_h$ \mike{in the simulation} are dominated by the magnetic field component parallel to the PIL, i.e., the shear component. \mike{This also matches observations}. This behavior is due to the change in inclination of the lower sections of magnetic field lines. They evolve from initially quasi-vertical geometries at the footpoints of large eruptive arcades to final inclined geometries at the footpoints of sheared flare-loops. This suggests a specific link with the flare loops. 

At a given position, the increase of the shear component is preceded by a decrease of the other component,  perpendicular to the PIL. These decreases are initially global, and they start right from the onset of the eruption. They are due to the CME that tends to stretch the field lines vertically all around it. This shows that the CME decreases the horizontal magnetic fields, instead of increasing them. 

Other regions of horizontal magnetic field variations develop at larger distances from the PIL. They are dominated by the evolution of the magnetic field component perpendicular to the PIL. Two extended field-decreasing regions form at the center of the magnetic flux concentrations (as observed). They correspond to the aforementioned CME-driven vertical stretching of coronal loops. One field-increasing region forms at the footpoints of $\Omega$-loops that gradually bend towards the photosphere during the eruption. This is due to a deflection of the CME from the vertical direction, caused by the asymmetry in the flux distribution in the photosphere. 

At a given position, there is a delay between the passage of the flare ribbon and the subsequent steady increase of the shear component. The delay becomes longer as time progresses. This implies a mechanism that involves an increasing response-time of the photosphere for longer times and larger distances from the PIL. 

Analysis of the coronal magnetic field shows that the photospheric area in which the shear component increases is merely the footprint of a broad and expanding coronal volume. This volume is the flare cusp which is filled by the ensemble of flare loops. It is surrounded by narrow electric currents which map down to the flare ribbons. The flare loops are sheared in the corona because they are formed by magnetic reconnection that transfers the magnetic shear from the pre-reconnection erupting-loops into them. 

Right after a flare loop has reconnected, its coronal apex is briefly accelerated downwards by the tension-driven reconnection-jet, while the edge of the cusp and therefore the flare ribbons move away. Then the flare loop gently contracts towards the non-moving photosphere that acts as a wall, and it eventually relaxes to a quasi-force-free state. The properties of the induction equation in ideal-MHD as applied to the geometry of the modeled flare loops readily account for the increasing photospheric shear component during this relaxation, and the increasing length of the flare loops accounts for increasing relaxation times after the ribbon has moved away. 

On one hand, the global magnetic energy monotonically decreases in the computational volume during the eruptive flare. But on the other hand, the horizontal magnetic field increases at all locations that eventually become part of the cusp. This increase dominates the weaker variation of the vertical field. This has two consequences in terms of energetics. Firstly it leads to a local increase in the magnetic-energy density in all these locations. 
\nnc{And secondly, it leads to an increase of the magnetic energy that is contained within the volume of cusp that contains the flare loops. These properties imply that the sole contraction of flare loops cannot be used a priori to account for the global magnetic energy decrease during an eruption. So the implosion conjecture does not easily apply to the contraction of flare loops. Instead, the latter is a mere consequence of the geometry of a reconnected loop, which involves strong field-line curvature and therefore a strong magnetic-tension force at their apex, that pulls the loop downwards ideally after reconnection has occurred.}

The usual observational proxy for the Lorentz force density \nna{(the alternative Lorentz force density)} displays a downward component that increases with time around the PIL. This could be interpreted as an evidence of strong forces acting on the photosphere. However, the expression for the proxy is dominated by its horizontal magnetic field term. Both maps are actually almost identical, albeit for their sign. \mike{The map of $\bm{j}\bm{\times}\bm{B}$, however, is significantly different.} It is concentrated along the cusp edge in the corona and the current ribbons in the photosphere, and it is much weaker. Also in the area around the PIL where the magnetic field increases, it displays very weak oscillations that are consistent with the asymptotic relaxation of flare loops towards the force-free state.

\nna{From a strictly mathematical point of view, it is possible to calculate the total Lorentz force for an active region (or a closed domain system) at a given time, based only on the vector magnetogram. The final result is a single value of the Lorentz force for one relatively isolated domain (e.g. active region) for every timestep. \mike{However, it is not possible to prepare a map of $\bm{j}\bm{\times}\bm{B}$ based only of photospheric vector magnetograms. Instead, researchers have used the alternative Lorentz force density. In practice, it could have been reliable, but our empirical findings based on the comparison of the maps of the $\bm{j\times B}$ and of the alternative Lorentz force density shows that its use is questionable.} Therefore, we argue that the conclusions derived from the map of the alternative Lorentz force density should be treated with caution.}

All these aforementioned findings imply that the photospheric horizontal magnetic-field increases during solar eruption map the footpoints of sheared flare loops, and that they are driven by the flare-reconnection in the corona, not by any other cause. This conclusion contradicts previous interpretations based on momentum conservation between the upward-moving CME and the underlying photosphere, or based on the implosion conjecture that involves joint energy and volume decreases not involved in flare reconnection and loop relaxation. 
\\
\\
We very much appreciate the valuable comments of the anonymous referee. This study benefited from financial support from the Programme National Soleil Terre (PNST) of the CNRS/INSU and the CNES, as well as from the Programme des Investissements d'Avenir (PIA) supervised by the ANR. The work of KB is funded by the LabEx Plas@Par which is driven by Sorbonne Universit\'e. The numerical simulation used in this work was executed on the HPC center MesoPSL which is financed by the project Equip@Meso as well as the R\'egion Ile-de-France. This work has been done within the LABEX Plas@par project, and received financial state aid managed by the Agence Nationale de la Recherche, as part of the programme "Investissements d'avenir" under the reference ANR-11-IDEX-0004-02.

\bibliography{pp1}

\end{document}